\documentclass[10pt,twocolumn,english]{IEEEtran}
\usepackage[T1]{fontenc}
\usepackage[]{graphicx}
\usepackage{wrapfig}
\usepackage{subfigure}
\usepackage{amsmath}
\usepackage{amssymb}
\usepackage{nicefrac}
\usepackage[]{algorithm}
\usepackage[]{algorithmic}
\usepackage{url}

\makeatletter

\usepackage{babel}

\newtheorem{thm}{Theorem}

\newcommand{\goodgap}{\hspace{\subfigtopskip}\hspace{\subfigbottomskip}}
\newcommand{\figminusgap}{\vspace{-2mm}}
\newcommand{\eqminusgap}{\vspace{-2mm}}

\begin{document}

\title{Exploiting Statistical Dependencies in Sparse Representations for Signal Recovery}

\author{Tomer Peleg, Yonina C. Eldar,~\IEEEmembership{Senior~Member,~IEEE}, and Michael Elad,~\IEEEmembership{Fellow,~IEEE}
\thanks{Copyright (c) 2012 IEEE. Personal use of this material is permitted. However, permission to use this material for any other purposes must be obtained from the IEEE by sending a request to pubs-permissions@ieee.org.}
\thanks{T. Peleg and Y. C. Eldar are with the Department of Electrical Engineering, Technion -- Israel Institute of Technology, Haifa 32000, Israel (e-mail: \{tomerfa@tx,yonina@ee\}.technion.ac.il). Y. C. Eldar is also a Visiting Professor at Stanford, CA. M. Elad is with the Computer Science Department, Technion -- Israel Institute of Technology, Haifa 32000, Israel (e-mail: elad@cs.technion.ac.il).}
\thanks{This work was supported in part by the Israel Science Foundation under Grants 170/10 and 599/08, and by the FP7-FET program, SMALL project (grant agreement no. 225913).}}

\makeatother

\maketitle
\vspace{-20pt}
\begin{abstract}
\noindent
Signal modeling lies at the core of numerous signal and image processing applications. A recent approach that has drawn considerable attention is sparse representation modeling, in which the signal is assumed to be generated as a combination of a few atoms from a given dictionary. In this work we consider a Bayesian setting and go beyond the classic assumption of independence between the atoms. The main goal of this paper is to introduce a statistical model that takes such dependencies into account and show how this model can be used for sparse signal recovery. We follow the suggestion of two recent works and assume that the sparsity pattern is modeled by a Boltzmann machine, a commonly used graphical model. For general dependency models, exact MAP and MMSE estimation of the sparse representation becomes computationally complex. To simplify the computations, we propose greedy approximations of the MAP and MMSE estimators. We then consider a special case in which exact MAP is feasible, by assuming that the dictionary is unitary and the dependency model corresponds to a certain sparse graph. Exploiting this structure, we develop an efficient message passing algorithm that recovers the underlying signal. When the model parameters defining the underlying graph are unknown, we suggest an algorithm that learns these parameters directly from the data, leading to an iterative scheme for adaptive sparse signal recovery. The effectiveness of our approach is demonstrated on real-life signals - patches of natural images - where we compare the denoising performance to that of previous recovery methods that do not exploit the statistical dependencies.
\end{abstract}

\begin{IEEEkeywords}
Sparse representations, signal synthesis, Bayesian estimation, MAP, MRF, Boltzmann machine, greedy pursuit, unitary dictionary, decomposable model, message passing, pseudo-likelihood, SESOP, image patches, denoising.
\end{IEEEkeywords}

\section{Introduction}

\label{sec:intro}

Signal modeling based on sparse representations is used in numerous signal and image processing applications, such as denoising, restoration, source separation, compression and sampling (for a comprehensive review see \cite{Bruckstein09}). In this model a signal $y$ is assumed to be generated as $y=Ax+e$, where $A$ is the dictionary (each of the columns in $A$ is typically referred to as an atom), $x$ is a sparse representation over this dictionary, and $e$ is additive white Gaussian noise. Throughout this work we shall assume that the dictionary is known and fixed, and our derivations consider both arbitrary and unitary dictionaries. Our goal is to recover the sparse representation $x$ from $y$, and by multiplying the outcome by $A$ we can achieve denoising. We will use the term "sparse signal recovery" (or just "sparse recovery" and "signal recovery") to describe the task of recovering $x$ from $y$ for both the case of arbitrary dictionaries and unitary ones.

Various works that are based on this model differ in their modeling of the sparse representation $x$. The classical approach to sparse recovery considers a deterministic sparse representation and signal recovery is formulated as a deterministic optimization problem. Some examples include greedy pursuit algorithms like orthogonal matching pursuit (OMP) and CoSaMP, and convex relaxations like basis pursuit denoising and the Dantzig selector (for comprehensive reviews see \cite{Bruckstein09,Tropp10}). Recent works \cite{La06,Lu08,Eldar09,Eldar10,Baraniuk10,Duarte11} suggested imposing additional assumptions on the support of $x$ (the sparsity pattern), which is still regarded deterministic. These works show that using structured sparsity models that go beyond simple sparsity can boost the performance of standard sparse recovery algorithms in many cases.

Two typical examples for such models are wavelet trees \cite{La06} and block-sparsity \cite{Eldar09,Eldar10}. The first accounts for the fact that the large wavelet coefficients of piecewise smooth signals and images tend to lie on a rooted, connected tree structure \cite{Mallat98}. The second model is based on the assumption that the signal exhibits special structure in the form of the nonzero coefficients occurring in clusters. This is a special case of a more general model, where the signal is assumed to lie in a union of subspaces \cite{Lu08,Eldar09}. Block-sparsity arises naturally in many setups, such as recovery of multi-band signals \cite{Mishali08,Mishali09} and the multiple measurement vector problem. However, there are many other setups in which sparse elements do not fit such simple models. In \cite{Baraniuk10} the authors propose a general framework for structured sparse recovery and demonstrate how both block-sparsity and wavelet trees can be merged into standard sparse recovery algorithms.

In many applications it can be difficult to provide one deterministic model that describes all signals of interest. For example, in the special case of wavelet trees it is well known that statistical models, such as hidden Markov trees (HMTs) \cite{Crouse98}, are more reliable than deterministic ones. Guided by this observation, it is natural to consider more general Bayesian modeling, in which the sparse representation is assumed to be a random vector. Many sparsity-favoring priors for the representation coefficients have been suggested in statistics, such as the Laplace prior, "spike-and-slab" (mixture of narrow and wide Gaussian distributions) and Student's $t$ distribution (for a comprehensive review see \cite{Seeger08}). However, the representation coefficients are typically assumed to be independent of each other.

Here we are interested in Bayesian modeling that takes into account not only the values of the representation coefficients, but also their sparsity pattern. In this framework sparsity is achieved by placing a prior distribution on the support, and the representation coefficients are modeled through a conditional distribution given the support. The most simple prior for the support assumes that the entries of the sparsity pattern are independent and identically distributed (i.i.d.) (see e.g. \cite{Schniter08}). However, in practice, atoms in the dictionary are often not used with the same frequency. To account for this behavior, we can relax the assumption that the entries are identically distributed and assign different probabilities to be turned "on" for each entry \cite{Turek11}.

Besides the modeling aspect, another key ingredient in Bayesian formulations is the design objective. Two popular techniques are maximum \emph{a posteriori} (MAP) and minimum mean square error (MMSE) estimators. Typically these estimators are computationally complex, so that they can only be approximated. For example, approximate MAP estimation can be performed using a wide range of inference methods, such as the relevance vector machine \cite{Tipping01} and Markov chain Monte Carlo (MCMC) \cite{Neal93}. Such estimators are derived in \cite{Seeger08,Ji08} based on sparsity-favoring priors on $x$ and approximate inference methods. In \cite{Schniter08,Elad09} approximate MMSE estimators are developed, based on an i.i.d prior on the support. Finally, in the special case of a square and unitary dictionary, assuming independent entries in the support and Gaussian coefficients, it is well known that the exact MAP and MMSE estimators can be easily computed \cite{Turek11,Michaeli12}.

Independence between the entries in the support can be a useful assumption, as it keeps the computational complexity low and the performance analysis simple. Nevertheless, this assumption can be quite restrictive and leads to loss of representation power. Real-life signals exhibit significant connections between the atoms in the dictionary used for their synthesis. For example, it is well known that when image patches are represented using the discrete cosine transform (DCT) or a wavelet transform, the locations of the large coefficients are strongly correlated. Recent works \cite{Duarte08,He09,Wolfe04,Garrigues07,Cevher08} have made attempts to go beyond the classic assumption of independence and suggested statistical models that take dependencies into account. The special case of wavelet trees is addressed in \cite{Duarte08}, where HMTs are merged into standard sparse recovery algorithms, in order to improve some of their stages and lead to more reliable recovery. Another statistical model designed to capture the tree structure for wavelet coefficients, was suggested in \cite{He09}. An approximate MAP estimator was developed there based on this model and MCMC inference.

Here we consider more general dependency models based on undirected graphs, which are also referred as Markov random fields (MRFs), and focus on the special case of a Boltzmann Machine (BM). To the best of our knowledge a BM structure for sparsity patterns was originally suggested in \cite{Wolfe04} in the context of Gabor coefficients. MCMC inference was used there for non-parametric Bayesian estimation. In \cite{Garrigues07} the authors also use a BM structure, which allows them to introduce the concept of interactions in a general sparse coding model. An approximate MAP estimator is then developed by means of Gibbs sampling and simulated annealing \cite{Neal93}. Finally, in \cite{Cevher08} a BM prior on the support is used in order to improve the CoSaMP algorithm. We will relate in more detail to the recent works which used BM-based modeling and emphasize differences between these works and our approach in Section \ref{sec:past works}.

The current paper is aimed at further exploring the BM-based model proposed in \cite{Wolfe04,Garrigues07,Cevher08}. Once we adopt the BM as a model for the support, several questions naturally arise: how to perform pursuit for finding the sparse representation, how to find the model parameters, and finally how to combine these tasks with dictionary learning. In this paper we address the first two questions. For pursuit we suggest using a greedy approach, which approximates the MAP and MMSE estimators and is suitable for any set of model parameters. We then make additional modeling assumptions, namely a unitary dictionary and a banded interaction matrix, and develop an efficient message passing algorithm for signal recovery which obtains the exact MAP estimate in this setup. For learning the Boltzmann parameters we suggest using a maximum pseudo-likelihood (MPL) approach and develop an efficient optimization algorithm for solving the MPL problem. Finally, we use a block-coordinate optimization approach to estimate both the sparse representations and the model parameters directly from the data. This results in an iterative scheme for adaptive sparse signal recovery.

The paper is organized as follows. In Section \ref{sec:motivation} we motivate the need for inserting probabilistic dependencies between elements in the support by considering sparse representations of image patches over a DCT dictionary. In Section \ref{sec:bkg} we introduce useful notions and tools from the graphical models field and explore the BM prior. Section \ref{sec:problem formulation} defines the signal model, along with the MAP and MMSE estimation problems. In Section \ref{sec:approximated MAP} we develop several greedy approximations of the MAP and MMSE estimators for the BM prior. We then present setups where the MAP problem can be solved exactly and develop an efficient algorithm for obtaining the exact solution in Section \ref{sec:exact MAP}. We explore the performance of these algorithms through synthetic experiments in Section \ref{sec:synthetic simulations}. Estimation of the model parameters and adaptive sparse signal recovery are addressed in Section \ref{sec:adaptive scheme}. The effectiveness of our approach is demonstrated on image patches in Section \ref{sec:image patches}. Finally, we discuss relations to past works in Section \ref{sec:past works}.

\section{Motivation}

\label{sec:motivation}

\begin{figure*}
\centering
\includegraphics[scale=0.3]{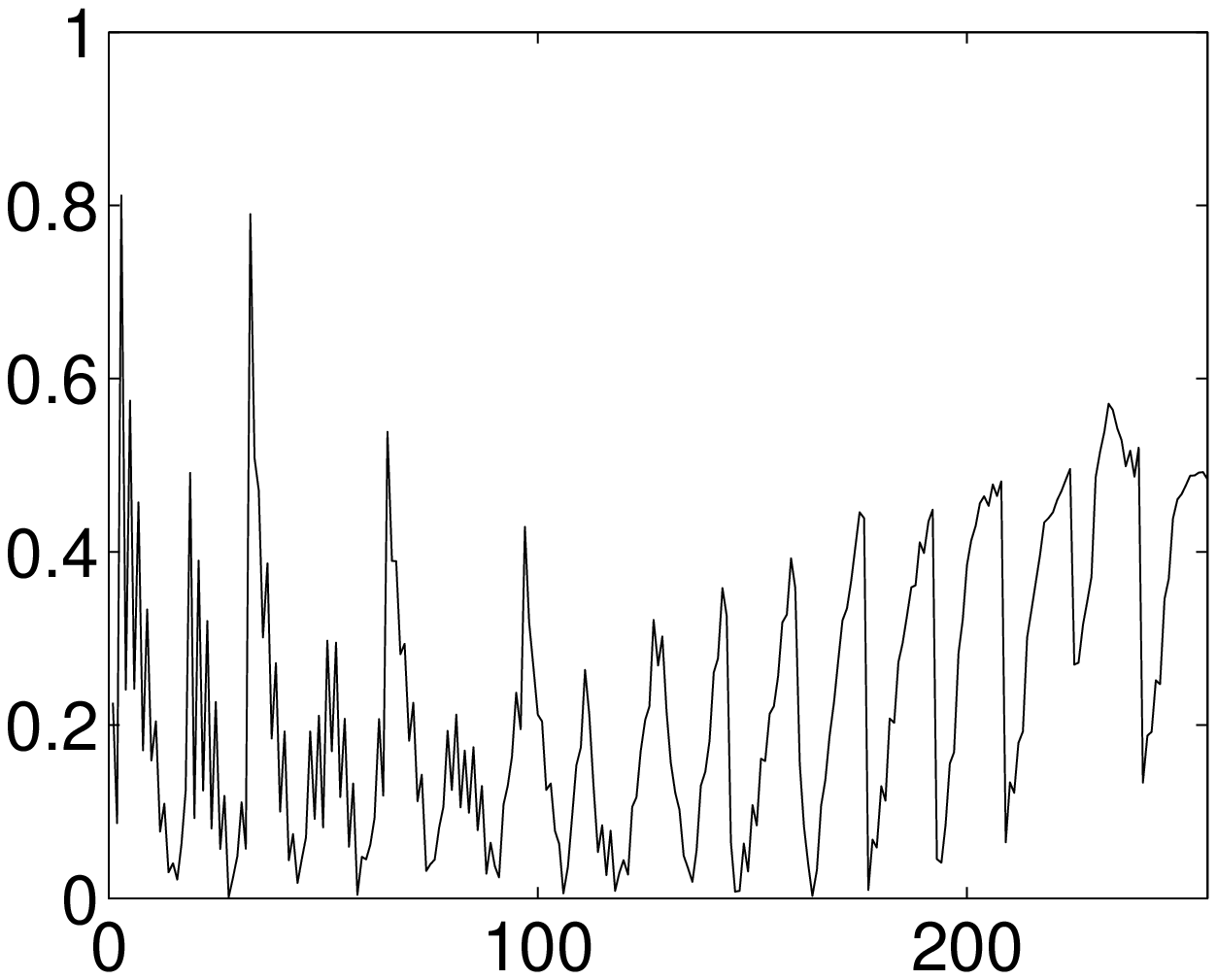}\goodgap
\includegraphics[scale=0.3]{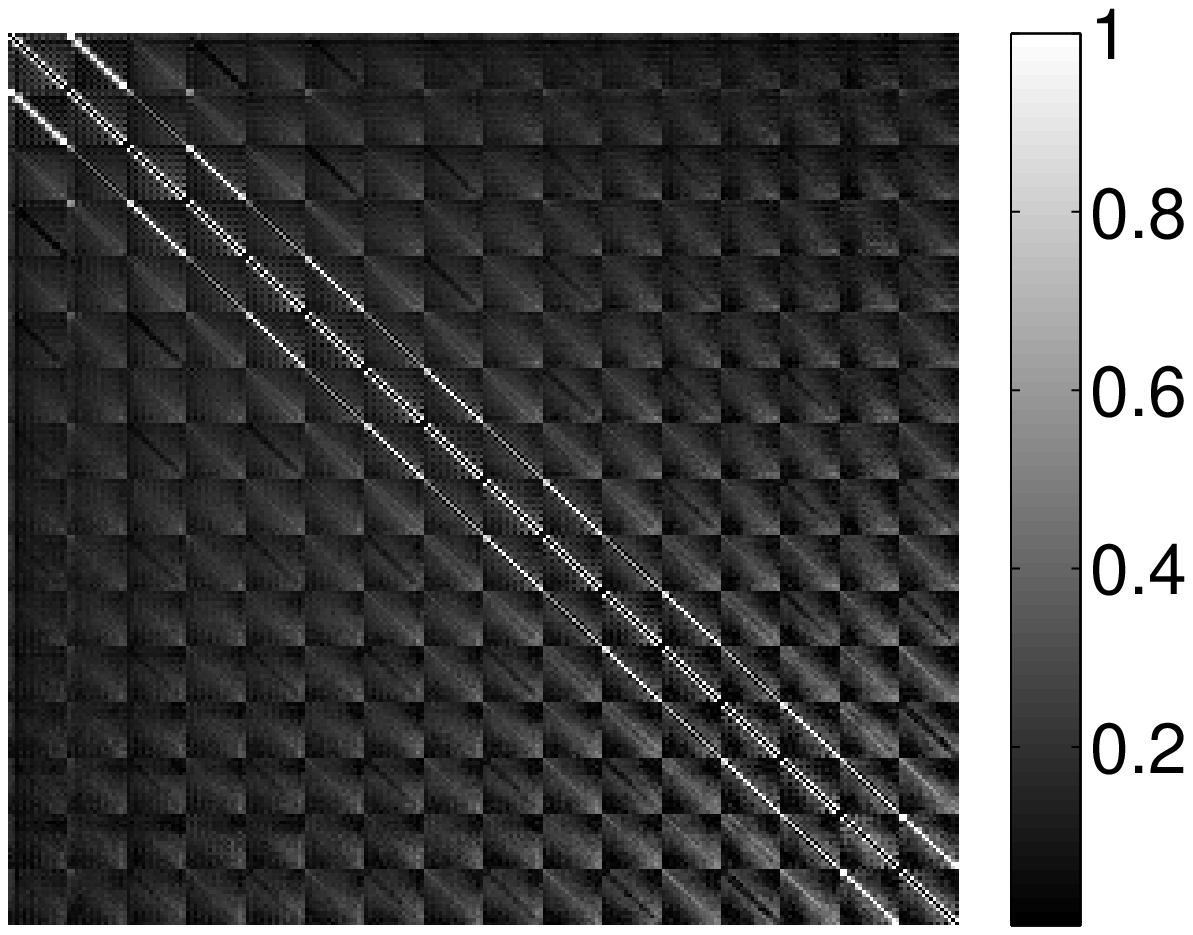}\goodgap
\includegraphics[scale=0.3]{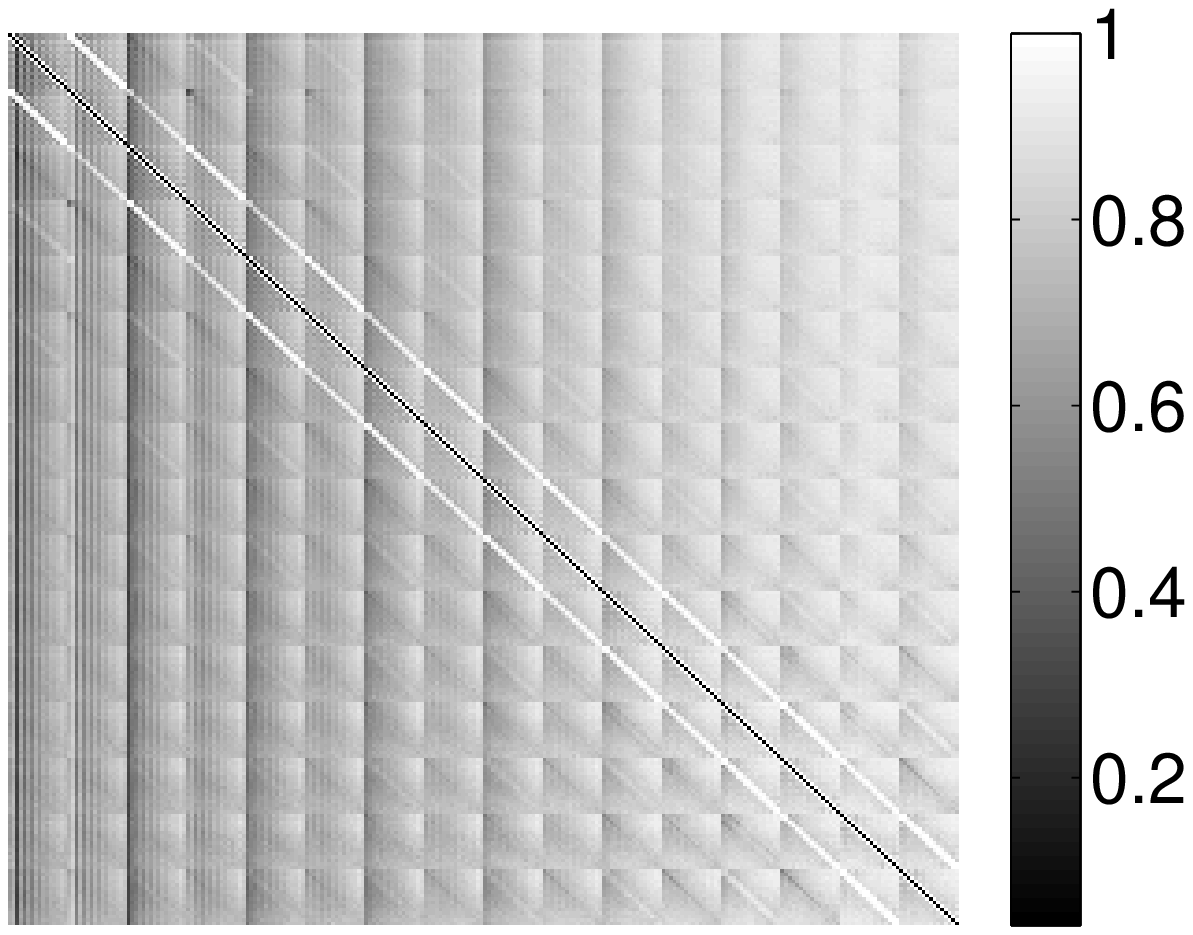}
\caption{Validity tests for several assumptions on the support vector: identical distributions, independency and block-sparsity. Left: A plot of $R$, Middle: An image of $U$, Right: An image of $V$.}
\label{fig:validity_tests}
\figminusgap
\end{figure*}

In this section we provide motivation for inserting probabilistic dependencies between elements in the support. We consider a set of $N=100,000$ patches of size $8$-by-$8$ that are extracted out of several noise-free natural images. For each patch, we perform a preliminary stage of DC removal by subtracting the average value of the patch, and then obtain sparse representations of these patches over an overcomplete DCT dictionary of size $64$-by-$256$ ($n$-by-$m$) using the OMP algorithm. We consider a model error of $\sigma=2$, so that OMP stops when the residual error falls below $\epsilon=\sqrt{n}\sigma=16$. We then compute the empirical marginal distributions for each of the dictionary atoms and for all pairs of atoms, namely we approximate $\Pr(S_i=1),~i=1,\ldots,m$ and $\Pr(S_i=1,S_j=1),~i=1,\ldots,m-1,~j>i$,
where $S$ is a binary vector of size $m$ and $S_i=1$ denotes that the $i$th atom is being used. The empirical conditional probability $\Pr(S_i=1|S_j=1)$ can then be computed as the ratio between $\Pr(S_i=1,S_j=1)$ and $\Pr(S_j=1)$.

We address several assumptions that are commonly used in the sparse recovery field and suggest validity tests for each of them. The first assumption is that the elements in the support vector are identically distributed, namely $\Pr(S_i=1)=p$ for all $i$, where $0\leq p\leq1$ is some constant. This assumption can be examined by comparing the marginal probabilities $\Pr(S_i=1)$ for each atom. The second assumption is independency between elements in the support. The independency assumption between atoms $i$ and $j$ implies that $\Pr(S_i=1|S_j=1)=\Pr(S_i=1)$. Therefore, we can test for independency by comparing the marginal and conditional probabilities for each pair of atoms. Next we turn to the block-sparsity assumption. Assuming that $i$ and $j$ are in the same cluster implies that the conditional probabilities $\Pr(S_i=1|S_j=1)$ and $\Pr(S_j=1|S_i=1)$ are near $1$.

To examine the validity of each of the above-mentioned assumptions, we compute the variables
\begin{align}
R_i=&\left|\log_{10}\left(\frac{\Pr(S_i=1)}{p}\right)\right|,~1\leq i\leq m\notag\\
U_{i,j}=&\left|\log_{10}\left(\frac{\Pr(S_i=1|S_j=1)}{\Pr(S_i=1)}+\delta\right)\right|,~1\leq i,j\leq m
\label{eq:validity}\\
V_{i,j}=&\left|\log_{10}\left(\Pr(S_i=1|S_j=1)+\delta\right)\right|,~1\leq i,j\leq m\notag
\eqminusgap
\end{align}
\noindent where $p$ denotes the average probability of an entry to be turned "on" , namely $p\triangleq\frac{1}{m}\sum_{l=1}^{m}\Pr(S_l=1)$, $R$ is a vector of size $m$ and $U,V$ are matrices of size $m$-by-$m$. We use $\delta=0.1$, so that for $\Pr(S_i=1|S_j=1)=0$ we get a value $1$ in $U_{i,j}$ and $V_{i,j}$ ($i$ and $j$ denote the row and column indices respectively). In each of the functions in (\ref{eq:validity}) a near-zero result implies that the corresponding assumption is valid; as we go further away from zero the validity of the assumption decreases. The logarithms are used to improve the visibility of the results.

The results are shown in Fig. \ref{fig:validity_tests}. On the left we plot the values in $R$. This plot demonstrates that the individual frequencies can be very far from the average one. Consequently, the DCT atoms are used with varying frequencies. The matrix $U$ is displayed in the middle. The black color, which corresponds to near-zero values, is dominant. This illustrates that the independency assumption is satisfactory for many pairs of DCT atoms. However, some pairs exhibit significant interactions (see the white diagonals near the main diagonal and the bright spots). The image on the right displays the matrix $V$, which is dominated by the white color, corresponding to near-one values. High values in the entries $V_{i,j}$ or $V_{j,i}$ indicate that it is not reasonable to assume that the corresponding atoms belong to the same cluster in a block-sparse model (regardless of the block sizes). Since this is the case for most pairs of DCT atoms, we conclude the block-sparsity approach does not capture the dependencies well in this example.

It is interesting to note that while the OMP algorithm reveals different frequencies of appearance for the atoms and significant correlations between pairs of atoms, it in fact makes no use of these properties. Therefore, it seems plausible that a stochastic model that will capture the different nature of each atom, as well as the important interactions between the atoms, can lead to improved performance. In this paper we will show how this can be accomplished in a flexible and adaptive manner. In Section \ref{sec:image patches} we will return to this very set of patches and show that the proposed model and methods do better service to this data.

\section{Background on Graphical Models}

\label{sec:bkg}

The main goal of this paper is using graphical models for representing statistical dependencies between elements in the sparsity pattern and developing efficient sparse recovery algorithms based on this modeling. In order to set the ground for the signal model and the recovery algorithms, we provide some necessary notions and methods from the vast literature on graphical models. We begin by presenting MRFs and explain how they can be used for describing statistical dependencies. We then focus on the BM, a widely used MRF, explore its properties and explain how it can serve as a useful and powerful prior on the sparsity pattern. For computational purposes we may want to relax the dependency model. One possible relaxation, which often reduces computational complexity and still bears considerable representation power, is decomposable models. Finally, we present a powerful method for probabilistic inference in decomposable models, coined belief propagation. Decomposability will be a modeling assumption in Section \ref{sec:exact MAP} and the algorithm we propose in Section \ref{subsec:message passing for exact MAP} will be based on belief propagation techniques.

\subsection{Representing Statistical Dependencies by MRFs}

\label{subsec:MRF}

\begin{figure}
\centering
\subfigure[Graph]{\includegraphics[scale=0.35]{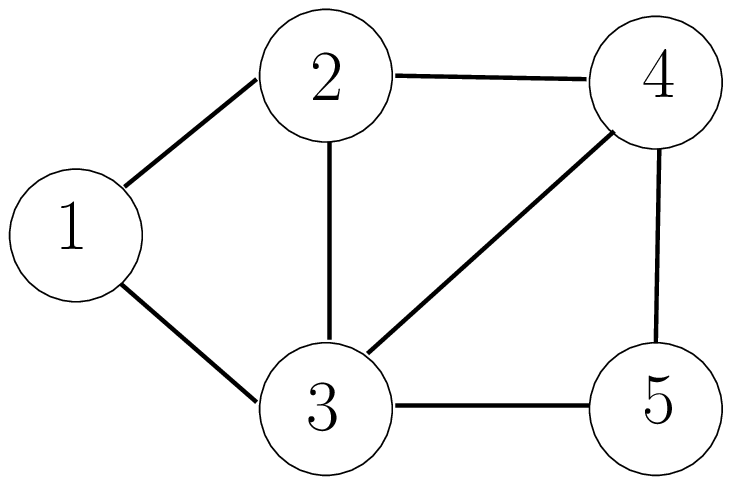}} \goodgap
\subfigure[Interaction matrix]{\includegraphics[scale=0.7]{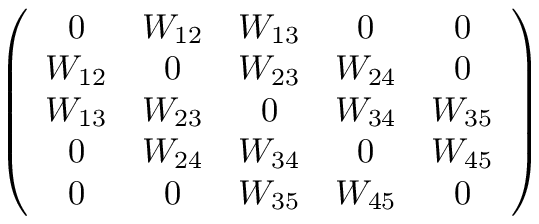}}
\caption{A simple dependency model for $5$ variables. This is a chordal graph with $3$ missing edges. The interaction matrix in the corresponding BM is banded.}
\label{fig:simple graph}
\figminusgap
\end{figure}

In this subsection we briefly review MRFs and how they can be used to represent statistical dependencies. This review is mainly based on \cite{Jordan04}. A graphical model is defined by its structural and parametric components. The structural component is the graph $G=(V,\varepsilon)$ where $V$ is a set of nodes (vertices) and $\varepsilon$ is a set of undirected edges (links between the nodes). In a graphical model there is a one-to-one mapping between nodes $\{1,2,\ldots,m\}$ and random variables $\left\{S_1,S_2,\ldots,S_m\right\}$. Let $S_A,S_B,S_C$ stand for three disjoint subsets of nodes. We say that $S_A$ is independent of $S_C$ given $S_B$ if $S_B$ separates $S_A$ from $S_C$, namely all paths between a node in $S_A$ and a node in $S_C$ pass via a node in $S_B$. Thus, simple graph separation is equivalent to conditional independence. The structure can be used to obtain all the global conditional independence relations of the probabilistic model. By "global" we mean that conditional independence holds for all variable assignments and does not depend on numerical specifications. For a visual demonstration see Fig. \ref{fig:simple graph}(a); using the above definition it easy to verify for example that $S_1$ is independent of $S_4,S_5$ given $S_2,S_3$.

Turning to the parametric component, note that the joint probability distribution is represented by a local parametrization. More specifically, we use a product of local nonnegative compatibility functions, which are referred to as potentials. The essence of locality becomes clearer if we define the notion of cliques. A clique is defined as a fully-connected subset of nodes in the graph. If $S_i$ and $S_j$ are linked, they appear together in a clique and thus we can achieve dependence between them by defining a potential function on that clique. The maximal cliques of a graph are the cliques that cannot be extended to include additional nodes without losing the property of being fully connected. Since all cliques are subsets of one or more maximal cliques, we can restrict ourselves to maximal cliques without loss of generality. For example, in Fig. \ref{fig:simple graph}(a) the maximal cliques are $C_1=\{1,2,3\}$, $C_2=\{2,3,4\}$ and $C_3=\{3,4,5\}$. To each maximal clique $C$ we assign a nonnegative potential $\Psi_C(S_C)$. The joint probability is then given as a product of these potentials, up to a normalization factor $Z$:
\begin{equation}
\Pr(S)\triangleq \frac{1}{Z}\prod_C~\Psi_C(S_C).
\label{eq:joint probability via potentials}
\end{equation}
\noindent If the potentials are taken from the exponential family, namely $\Psi_C(S_C)=\exp\left\{-E_C(S_C)\right\}$, then $\Pr(S)=\frac{1}{Z}\exp\{-E(S)\}$, where $E(S)=\sum_C~E_C(S_C)$ is the energy of the system.

\subsection{The Boltzmann Machine}

\label{subsec:Boltzmann}

In this subsection we focus on the BM, a widely used MRF. We are about to show that this can serve as a useful and powerful prior on the sparsity pattern. The BM distribution is given by:
\begin{equation}
\Pr(S)=\frac{1}{Z}\exp\left(b^TS+\frac{1}{2}S^TWS\right),
\label{eq:BM prior}
\end{equation}
\noindent where $S$ is a binary vector of size $m$ with values in $\{-1,1\}^m$, $W$ is symmetric and $Z$ is a partition function of the Boltzmann parameters $W,b$ that normalizes the distribution. We can further assume that the entries on the main diagonal of $W$ are zero, since they contribute a constant to the function $S^TWS$. In this work the BM will be used as a prior on the support of a sparse representation: $S_i=1$ implies that the $i$th atom is used for the representation, whereas for $S_i=-1$ this atom is not used.

The BM is a special case of the exponential family with an energy function $E(S)=-b^TS-\frac{1}{2}S^TWS$. The BM distribution can be easily represented by an MRF - a bias $b_i$ is associated with a node $i$ and a nonzero entry $W_{ij}$ in the interaction matrix results in an edge connecting nodes $i$ and $j$ with the specified weight. Consequently, the zero entries in $W$ have the simple interpretation of missing edges in the corresponding undirected graph. This means that the sparsity pattern of $W$ is directly linked to the sparsity of the graph structure. From graph separation we get that if $W_{ij}=0$ then $S_i$ and $S_j$ are statistically independent given all their neighbors $\{S_l\}_{l\in~N(i)\cup N(j),~l\neq i,j}$. For example, if the matrix $W$ corresponds to the undirected graph that appears in Fig. \ref{fig:simple graph}(a) then $W_{14}=W_{15}=W_{25}=0$. This matrix is shown in Fig. \ref{fig:simple graph}(b).

The maximal cliques in the BM are denoted by $C_1,\ldots,C_P$ and we would like to assign potential functions $\left\{\Psi_{C_i}\left(S_{C_i}\right)\right\}_{i=1}^{P}$ to these cliques that will satisfy the requirement $\exp\left(b^TS+\frac{1}{2}S^TWS\right)=\prod_{i=1}^{P}\Psi_{C_i}\left(S_{C_i}\right)$. One possible choice is to assign each of the terms in $E(S)$ using a pre-specified order of the cliques: $b_iS_i$ is assigned to the clique that consists of $S_i$ and appears last in the order and a non-zero term $W_{ij}S_iS_j$ is assigned to the clique that consists of $S_i,S_j$ and appears last in the order.

Next, we turn to explore the intuitive meaning of the Boltzmann parameters. In the simple case of $W=0$, the BM distribution becomes $\Pr(S)=\frac{1}{Z}\prod_{i=1}^m \exp\left(b_iS_i\right)$. Consequently, $\left\{S_i\right\}_{i=1}^m$ are statistically independent. Using straightforward computations we get $\Pr(S_i=-1)=\exp(-2b_i)\Pr(S_i=1)$ for $i=1,\ldots,m$. Since $\Pr(S_i=-1)+\Pr(S_i=1)=1$, $S_i$ has the following marginal probability to be turned "on":
\begin{equation}
p_i\triangleq\Pr(S_i=1)=\frac{1}{1+\exp(-2b_i)},~1\leq i\leq m.
\label{eq:marginals independence}
\end{equation}
\noindent When $W$ is nonzero, (\ref{eq:marginals independence}) no longer holds. However, the simple intuition that $S_i$ tends to be turned "off" as $b_i$ becomes more negative, remains true.

We would now like to understand how to describe correlations between elements in $S$. To this end we focus on the simple case of a matrix $W$ of size $2$-by-$2$, consisting of one parameter $W_{12}$, and provide an exact analysis for this setup. In order to simplify notations, from now on we use $p_{i|j}(u|v)$ to denote $\Pr(S_i=u|S_j=v)$. Using these notations we can write down the following relation for the simple case of a pair of nodes:
\begin{equation}
p_1=p_{1|2}(1|1)p_2+p_{1|2}(1|-1)(1-p_2),
\label{eq:marginal for a pair}
\eqminusgap
\end{equation}
\noindent where
\begin{eqnarray} &&p_{1|2}(1|1)=\frac{1}{1+\exp(-2b_1-2W_{12})} \notag\\
&&p_{1|2}(1|-1)=\frac{1}{1+\exp(-2b_1+2W_{12})}.
\label{eq:conditionals for a pair}
\end{eqnarray}
\noindent From (\ref{eq:marginal for a pair}) we see that $p_1$ is a convex combination of $p_{1|2}(1|-1)$ and $p_{1|2}(1|1)$. Hence, for $W_{12}>0$ we have $p_{1|2}(1|-1)<p_1<p_{1|2}(1|1)$ and for $W_{12}<0$ we have $p_{1|2}(1|1)<p_1<p_{1|2}(1|-1)$.

For a general matrix $W$ these relations are no longer strictly accurate. However, they serve as useful rules of thumb: for an "excitatory" interaction ($W_{ij}>0$) $S_i$ and $S_j$ tend to be turned "on" ("off") together, and for an "inhibitory" interaction ($W_{ij}<0$) $S_i$ and $S_j$ tend to be in opposite states. The intuition into the Boltzmann parameters provides some guidelines as to how the BM prior can be used for sparse representations. If the values of the biases in the vector $b$ are negative "enough" and there are few strong excitatory interactions, then the mean cardinality of the support tends to be small. This reveals some of the power of the BM as a prior on the support in the signal model. It can achieve sparsity and at the same time capture statistical dependencies and independencies in the sparsity pattern.

To conclude this section, note that standard sparsity-favoring models can be obtained as special cases of the BM model. For $W=0$ and $b_i=\frac{1}{2}\ln\left(\frac{p}{1-p}\right)$ for all $i$, which correspond to an i.i.d. prior, the cardinality $k$ has a Binomial distribution, namely $k\sim Bin(p,m)$. For a low value of $p$ the cardinalities are typically much smaller than $m$, so that plain sparsity is achieved. BM can also describe a block-sparsity structure: Assuming that the first $k_1$ entries in $S$ correspond to the first block, the next $k_2$ to the second block, etc., the interaction matrix $W$ should be block-diagonal with "large" and positive entries within each block. The entries in $b$ should be chosen as mentioned above to encourage sparsity.

\subsection{Decomposable Graphical Models}

\label{subsec:decomposable models}

We now consider decomposability in graphical models \cite{Jordan04,Wiesel10}. A triplet $\{A,B,C\}$ of disjoint subsets of nodes is a decomposition of a graph if its union covers all the set $V$, $B$ separates $A$ from $C$ and $B$ is fully-connected. It follows that a graphical model is regarded as decomposable if it can be recursively decomposed into its maximal cliques, where the separators are the intersections between the cliques. It is well known that a decomposable graph is necessarily chordal \cite{Lauritzen96}. This means that each of its cycles of four or more nodes has a chord, which is an edge joining two nodes that are not adjacent in the cycle. Consequently, for a given MRF we can apply a simple graphical test to verify that it is decomposable.

In Section \ref{sec:exact MAP} we consider decomposable BMs. This assumption implies that the matrix $W$ corresponds to a chordal graph. We now provide some important examples for decomposable graphical models and their corresponding interaction matrices. Note that a graph which contains no cycles of length four is obviously chordal as it satisfies the required property in a trivial sense. It follows that a graph with no edges, a graph consisting of non-overlapping cliques and a tree are all chordal. The first example is the most trivial chordal graph and corresponds to $W=0$. The second corresponds to a block-diagonal matrix and as we explained in Section \ref{subsec:Boltzmann} it can describe a block-sparsity structure. Tree structures are widely used in applications that are based on a multiscale framework. A visual demonstration of the corresponding matrix is shown in \cite{Wiesel10}.

Another common decomposable model corresponds to a banded interaction matrix. In an $L$th order banded matrix only the $2L+1$ principal diagonals consist of nonzero elements. Assuming that the main diagonal is set to zero, we have that there can be at most $(2m-(L+1))L$ nonzero entries in an $L$th order banded $W$, instead of $m^2-m$ nonzeros as in a general interaction matrix. Consequently, the sparsity ratio of $W$ is of order $\nicefrac{L}{m}$. This matrix corresponds to a chordal graph with cliques $C_{i}=\left\{S_{i},\ldots,S_{i+L}\right\},~i=1,\ldots,m-L$. For example, the matrix in Fig. \ref{fig:simple graph}(b) is a second order banded matrix of size $5$-by-$5$. This matrix corresponds to a chordal graph (see Fig. \ref{fig:simple graph}(a)) with three cliques.

\begin{figure}
\centering
\includegraphics[scale=0.4]{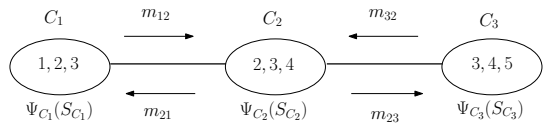}
\caption{A clique tree which is constructed for the graph that appears in Fig. \ref{fig:simple graph}. In this case the clique tree takes the form of a simple chain of size $3$. Potential functions are defined for each of the cliques and exact probabilistic inference is performed by message passing.}
\label{fig:simple clique tree}
\figminusgap
\end{figure}

Chordal graphs serve as a natural extension to trees. It is well known \cite{Jordan04} that the cliques of a chordal graph can be arranged in a clique tree, which is called a junction tree. In a junction tree $T$ each clique serves as a vertex and any two cliques containing a node $v$ are either adjacent in $T$ or connected by a path made entirely of cliques containing $v$. For a visual demonstration see Fig. \ref{fig:simple clique tree}, where a clique tree is constructed for the chordal graph of Fig. \ref{fig:simple graph}(a). In this case where the interaction matrix is banded, the clique tree is simply a chain. It can easily be verified that this is in fact true for a banded interaction matrix of any order.

We now turn to describe belief propagation, a powerful method for probabilistic inference tasks like computation of single node marginal distributions and finding the most probable configuration. Exact probabilistic inference can become computationally infeasible for general dependency models as it requires a summation or maximization over all possible configurations of the variables. For example, in a general graphical model with $m$ binary variables the complexity of exact inference grows exponentially with $m$. However, when the graph structure is sparse, one can often exploit the sparsity in order to reduce this complexity. The inference tasks mentioned above can often be performed efficiently using belief propagation techniques \cite{Jordan04}. More specifically, in a decomposable MRF exact inference takes the form of a message passing algorithm, where intermediate factors are sent as messages along the edges of the junction tree (see for example the messages passed along the chain in Fig. \ref{fig:simple clique tree}). For more details on message passing see \cite{Jordan04}.

The complexity of exact inference via message passing strongly depends on the tree-width of the graph. In a decomposable model this is defined as the size of the largest maximal clique minus one. For example, in the special case of a BM with an $L$th order banded $W$ we have that the tree-width is $L$. We can conclude that for a decomposable model there is an obvious tradeoff between computational complexity and representation power. For example, in the special case of an $L$th order interaction matrix the computational complexity of exact inference decreases with $L$, but at the same time the graphical model captures fewer interactions. Nevertheless, decomposable models can serve as a useful relaxation for a general dependency model, as they can achieve a substantial decrease in the complexity of exact inference, while still capturing the significant interactions.

\section{BM Generative Model}

\label{sec:problem formulation}

In this section we use the BM for constructing a stochastic generative signal model. We consider a signal $y$ which is modeled as $y=Ax+e$, where $A$ is the dictionary of size $n$-by-$m$, $x$ is a sparse representation over this dictionary and $e$ is additive white Gaussian noise with variance $\sigma_e^2$. We denote the sparsity pattern by $S\in\{-1,1\}^m$, where $S_i=1$ implies that the index $i$ belongs to the support of $x$ and $S_i=-1$ implies that $x_i=0$. The nonzero coefficients of $x$ are denoted by $x_s$, where $s$ is the support of $x$. Following \cite{Garrigues07} we consider a BM prior for $S$ and a Gaussian distribution with zero mean and variance $\sigma_{x,i}^2$ for each nonzero representation coefficient $x_i$. Note that the variances of the non-zero representation coefficients are atom-dependent. It follows that the conditional distribution of $x_s$ given the support $s$ is
\begin{equation}
\Pr(x_s|s)=\frac{1}{\det\left(2\pi\Sigma_s\right)^{1/2}}
\exp\left\{-\frac{1}{2}x_s^T\Sigma_s^{-1}x_s\right\} \label{eq:conditional for x}
\end{equation}
\noindent where $\Sigma_s$ is a $k\times k$ diagonal matrix with diagonal elements $(\Sigma_s)_{i,i}=\sigma_{x,s_i}^2$ and $k$ is the cardinality of the support $s$. Using the assumption on the noise we can also write down the conditional distribution for the signal $y$ given its sparse representation:
\begin{equation}
\Pr(y|x_s,s)=\frac{1}{\left(2\pi\sigma_e^2\right)^{n/2}}
\exp\left\{-\frac{1}{2\sigma_e^2}\|y-A_sx_s\|_2^2\right\}. \label{eq:conditional for y}
\end{equation}

The proposed generative model combines a discrete distribution for $S$ and continuous distributions for $x$ given $S$ and $y$ given $x$, so that computations of posterior distributions should be handled carefully. Notice that an empty support $s$ necessarily implies $x=0$, so that $\Pr(x=0)$ is a discrete distribution (it equals $\Pr(S=-\textbf{1}^{m\times1})$). However, for nonzero vectors $v$ we have that $\Pr(x=v)$ is a continuous distribution. Using Bayes' law we can deduce that just like $\Pr(x)$, the posterior $\Pr(x|y)$ is a mixture of a discrete distribution for $x=0$ and a continuous distribution for all nonzero values of $x$. Our goal is to recover $x$ given $y$. However, from the above discussion we have that given $y$, the representation vector $x$ equals zero with a nonzero probability, whereas for any nonzero vector $v$ the event $x=v$ occurs with probability zero. It follows that the MAP estimator for $x$ given $y$ leads to the trivial solution $x=0$, rendering it useless.

The distribution $\Pr(s|y)$ however is a discrete one. Therefore, we suggest to first perform MAP estimation of $s$ given $y$ and then proceed with MAP estimation of $x$ given $y$ and the estimated support $\hat{s}$ \cite{Turek11}. This suggestion aligns with previous approaches in the sparse recovery field. In fact, standard algorithms for sparse recovery, such as OMP, take a similar approach - they first obtain an estimate for the support which minimizes the residual error and then rely on this estimate for signal reconstruction. Indeed, even the celebrated $l_1$-norm minimization approach is often used as a means to find the support, followed by a least-squares step for finding the final representation values (this is known as debiasing).

We begin by developing an expression for $\Pr(y|s)$ by integrating over all possible values of $x_s\in\mathbb{R}^k$:
\small
\begin{align}
&\Pr(y|s)=\int_{x_s\in\mathbb{R}^k}~\Pr(y|x_s,s)\Pr(x_s|s)dx_s\notag\\
&=C\frac{1}{\det\left(\frac{1}{\sigma_e^2}A_s^TA_s\Sigma_s+I\right)^{1/2}}\exp\left\{\frac{1}{2\sigma_e^2}
y^TA_sQ_s^{-1}A_s^Ty\right\}\label{eq:y given s}
\eqminusgap
\end{align}
\normalsize
\noindent where $C=\nicefrac{1}{\left(2\pi\sigma_e^2\right)^{n/2}}\exp\left\{-\frac{1}{2\sigma_e^2}\|y\|_2^2\right\}$ is a constant and $Q_s=A_s^TA_s+\sigma_e^2\Sigma_s^{-1}$. This leads to the following estimator for the support:
\begin{align}
\hat{s}_{_{MAP}}=&\underset{s\in\Omega}{\textnormal{argmax}}\Pr(s|y)=\underset{s\in\Omega}{\textnormal{argmax}}
\Pr(y|s)\Pr(s)\notag\\
=&\underset{s\in\Omega}{\textnormal{argmax}}\frac{1}{2\sigma_e^2}y^TA_sQ_s^{-1}A_s^Ty-\label{eq:MAP for s}\\
&\frac{1}{2}\ln\left(\det\left(Q_s\right)\right)+\frac{1}{2}S^TWS+\left(b-\frac{1}{4}v\right)^TS\notag
\end{align}
\noindent where $v_i=\ln\left(\nicefrac{\sigma_{x,i}^2}{\sigma_e^2}\right)$ and $S$ depends on $s$ through $S_i=2\cdot\textbf{1}[i\in s]-1$ for all $i$, with $\textbf{1}[\cdot]$ denoting the indicator function. The feasible set $\Omega$ denotes all $2^m$ possible supports. In terms of $S$, this is the set of all vectors satisfying $S_i^2=1$ for all $i$. Note that for an empty support the two first terms in (\ref{eq:MAP for s}) vanish. Once we have an estimate $\hat{s}=\hat{s}_{MAP}$ of the support, we can compute a MAP estimator of $x$ using the same formula as in the oracle estimator (see \cite{Turek11}):
\begin{equation}
\hat{x}_{\hat{s}_{_{MAP}}}=\underset{x_{\hat{s}}\in\mathbb{R}^k}{\textnormal{argmax}}\Pr\left(x_{\hat{s}}|y,\hat{s}\right)=Q_{\hat{s}}^{-1}A_{\hat{s}}^Ty.
\label{eq:MAP for x}
\end{equation}

We now turn to MMSE estimation of $x$ given $y$. Here we have:
\begin{equation}
\hat{x}_{MMSE}=E[x|y]=\sum_{s\in\Omega}\Pr(s|y)E[x|y,s],
\label{eq:MMSE for x}
\end{equation}
\noindent where $E[x|y,s]$ equals $\underset{x}{\textnormal{argmax}}\Pr\left(x|y,s\right)$ \cite{Turek11} (this is not true in general, but rather for the specific distribution considered here) and is computed using the oracle formula: $E[x_s|y,s]=Q_s^{-1}A_s^Ty$ for the on support entries (and the off support entries are set to zero).

In the sequel we first focus on the case where all model parameters - the Boltzmann parameters $W,b$, the variances $\left\{\sigma_{x,i}^2\right\}_{i=1}^m$, the dictionary $A$ and the noise variances $\sigma_e^2$ are known. For a general dictionary $A$ and an arbitrary symmetric interaction matrix $W$ the exact MAP and MMSE estimators require an exhaustive search or sum over all $2^m$ possible supports. To overcome the infeasibility of the combinatorial search or sum, we consider two approaches. In the first, developed in Section \ref{sec:approximated MAP}, we approximate the MAP and MMSE estimators using greedy methods. An alternative strategy is to make additional assumptions on the model parameters, namely on $A$ and $W$, that will make exact estimation feasible. This approach is addressed in Section \ref{sec:exact MAP}, where we consider unitary dictionaries $A$ and decomposable BMs. The more practical setup where the model parameters are also unknown is considered in Section \ref{sec:adaptive scheme}, for which we derive efficient methods for estimating both the sparse representations and the model parameters from a set of signals.

\section{Greedy Pursuit for Approximate MAP and MMSE estimation}

\label{sec:approximated MAP}

Throughout this section we assume an arbitrary dictionary and an arbitrary symmetric interaction matrix and make use of the BM-based generative model to solve a fundamental sparse coding problem - finding the sparse representation of a signal from noisy observations. As we have seen in the previous section, exact MAP and MMSE estimation in this setup require an exhaustive search or sum over all $2^m$ possible supports. To simplify the computations, we propose using a greedy approach. In this section we suggest three greedy pursuit algorithms for our model-based sparse recovery problem. The two first algorithms are OMP-like and thresholding-like pursuits which approximate the MAP estimate of the support $s$ given the signal $y$. The third pursuit method is a randomized version of the proposed OMP-like algorithm (similar to the rand-OMP method \cite{Elad09}), which approximates the MMSE estimate of the representation vector $x$ given the signal $y$.

\subsection{OMP-like MAP}

\label{subsec:OMP-like}

We begin with the OMP-like algorithm and explain its core idea. Our goal is to estimate the support which achieves the maximal value of the posterior probability $\Pr(S|y)$. This means that our objective function is the one that appears in (\ref{eq:MAP for s}). We start with an empty support, which means that $\left\{S_i\right\}_{i=1}^m$ are all $-1$. At the first iteration, we check each of the $m$ possible elements that can be added to the empty support and evaluate the term in (\ref{eq:MAP for s}). The entry $i_*$ leading to the largest value is chosen and thus $S_{i_*}$ is set to be $+1$. Given the updated support, we proceed exactly in the same manner. In every iteration we consider all the remaining inactive elements and choose the one that leads to the maximal value in (\ref{eq:MAP for s}) when added to the previously set support. The algorithm stops when the value of (\ref{eq:MAP for s}) is decreased for every additional item in the support.

In each iteration only one entry in $S$ changes - from $-1$ to $1$. This can be used to simplify some of the terms that appear in (\ref{eq:MAP for s}):
\begin{align}
&\frac{1}{2}S^TWS=\frac{1}{2}\sum_{i,j}W_{ij}S_iS_j=C_1+2\sum_jW_{ij}S_j\notag\\
&b^TS=\sum_{i=1}^mb_iS_i=C_2+2b_i\label{eq:greedy one iteration}\\
&\sum_{i=1}^m\ln\left(\nicefrac{\sigma_{x,i}^2}{\sigma_e^2}\right)S_i=C_3+2\ln\left(\sigma_{x,i}^2\right)\notag
\end{align}
\noindent where $C_1,C_2,C_3$ are constants that will not be needed in our derivation. Consequently, in each iteration it is sufficient to find an index $i$ (out of the remaining inactive indices) that maximizes the following expression:
\begin{align}
Val(i)=&\frac{1}{2\sigma_e^2}y^TA_{s^k}Q_{s^k}^{-1}A_{s^k}^Ty-
\frac{1}{2}\ln\left(\left|\det\left(Q_{s^k}\right)\right|\right)+\notag\\
&2W_i^TS^k+2b_i-\frac{1}{2}\ln\left(\sigma_{x,i}^{2}\right)
\label{eq:Val}
\end{align}
\noindent where $s_k$ is the support estimated in iteration $k-1$ with the entry $i$ added to it, $Q_{s^k}=A_{s^k}^TA_{s^k}+\sigma_e^2\Sigma_{s^k}^{-1}$ and $W_i^T$ is the $i$th row of $W$. A pseudo-code for the proposed OMP-like algorithm is given in Algorithm \ref{alg:greedy MAP}.

\begin{algorithm}
\caption{Greedy OMP-like algorithm for approximating the MAP estimator of (\ref{eq:MAP for s})}
\label{alg:greedy MAP}
\begin{algorithmic}
\REQUIRE{Noisy observations $y\in\mathbb{R}^n$ and model parameters $W,b,\left\{\sigma_{x,i}\right\}_{i=1}^m,A,\sigma_e$.}
\ENSURE{A recovery $\hat{s}_{\textnormal{MAP}}$ for the support.}
\STATE $s_*^0=\emptyset,~S_*^0=-\textbf{1}^{m\times1}$
\STATE $k=1$
\REPEAT
\FOR {$i\notin s_{*}^{k-1}$}
\STATE $s^k=s_*^{k-1}\cup i$
\STATE $S^k[j]=\begin{cases} S_*^{k-1}[j] &,\quad j\neq i\\
1 &,\quad j=i\end{cases}$
\STATE Evaluate $Val(i)$ using $(\ref{eq:Val})$.
\ENDFOR
\STATE $i_*=\textnormal{argmax}_i\left\{Val(i)\right\}$
\STATE $s_*^k=s_*^{k-1}\cup i_*,~S_*^k[j]=\begin{cases} S_*^{k-1}[j] &,\quad j\neq i_*\\
1 &,\quad j=i_*\end{cases}$
\STATE $k=k+1$
\UNTIL{$\Pr\left(s_*^k|y\right)<\Pr\left(s_*^{k-1}|y\right)$}
\STATE \textbf{Return:} $\hat{s}_{\textnormal{MAP}}=s_{*}^{k-1}$
\end{algorithmic}
\end{algorithm}

We now provide some intuition for the expressions in (\ref{eq:Val}). The term
$y^TA_{s^k}Q_{s^k}^{-1}A_{s^k}^Ty$ is equivalent to the residual error $\left\|r^k\right\|_2^2$, where $r^k=y-A_{s^k}\left(A_{s^k}^TA_{s^k}\right)^{-1}A_{s^k}^Ty$ is the residual with respect to the signal. To see this, notice that the following relation holds:
\begin{equation}
\left\|r^k\right\|_2^2=\left\|y\right\|_2^2-y^TA_{s^k}\left(A_{s^k}^TA_{s^k}\right)^{-1}A_{s^k}^Ty.
\label{eq:residual error}
\end{equation}
\noindent Using the definition of $Q_{s^k}$ it can be easily verified that the two terms take a similar form, up to a regularization factor in the pseudoinverse of $A_{s^k}$. Next, we turn to the terms $W_i^TS^k$ and $b_i$. The first corresponds to the sum of interactions between the $i$th atom and the rest of the atoms which arise from turning it on (the rest remain unchanged). The second term is the separate bias for the $i$th atom. As the sum of interactions and the separate bias become larger, using the $i$th atom for the representation leads to an increase in the objective function. Consequently, the total objective of (\ref{eq:Val}) takes into consideration both the residual error with respect to the signal and the prior on the support. This can lead to improved performance over standard pursuit algorithms like OMP and CoSaMP, which are aimed at minimizing the residual error alone.

\subsection{Thresholding-like MAP}

\label{subsec:THR-like}

To simplify computations, we can consider a thresholding-like version of Algorithm \ref{alg:greedy MAP}. Again we start with an empty support and compute $Val(i)$ using $(\ref{eq:Val})$ for $i=1,\ldots,m$, just as we do in the first iteration of Algorithm \ref{alg:greedy MAP}. We then sort the indices according to $Val(i)$ in a descending order and consider $m$ candidate supports for solving the MAP estimation problem, where the $k$th candidate consists of the first $k$ elements in the above order. Among these supports we choose the one that maximizes the posterior probability $\Pr(S|y)$. A pseudo-code for the proposed thresholding-like algorithm is given in Algorithm \ref{alg:THR-like MAP}.

\begin{algorithm}
\caption{Greedy thresholding-like algorithm for approximating the MAP estimator of (\ref{eq:MAP for s})}
\label{alg:THR-like MAP}
\begin{algorithmic}
\REQUIRE{Noisy observations $y\in\mathbb{R}^n$ and model parameters $W,b,\left\{\sigma_{x,i}\right\}_{i=1}^m,A,\sigma_e$.}
\ENSURE{A recovery $\hat{s}_{\textnormal{MAP}}$ for the support.}
\FOR {$i\in\{1,\ldots,m\}$}
\STATE $s={i}$
\STATE $S[j]=\begin{cases} -1 &,\quad j\neq i\\
1 &,\quad j=i\end{cases}$
\STATE Evaluate $Val(i)$ using $(\ref{eq:Val})$.
\ENDFOR
\STATE Sort $Val(i)$ in a descending order and arrange the indices $1,\ldots,m$ according to this order.
\FOR {$k\in\{1,\ldots,m\}$}
\STATE Set $s^{(k)}$ to include the first $k$ elements in above order.
\STATE Compute $\Pr\left(s^{(k)}|y\right)$.
\ENDFOR
\STATE $k_*=\textnormal{argmax}_k\left\{\Pr\left(s^{(k)}|y\right)\right\}$
\STATE \textbf{Return:} $\hat{s}_{\textnormal{MAP}}=s^{(k_*)}$
\end{algorithmic}
\end{algorithm}
\vspace{-20pt}

\subsection{Random OMP-like MMSE}

\label{subsec:rand OMP-like}

Another alternative is using a randomized version of Algorithm \ref{alg:greedy MAP} which approximates the MMSE estimate. The algorithmic framework remains the same as before, except for two changes. First, instead of adding to the support the element that maximizes $Val(i)$ in each iteration, we make a random choice with probabilities $\frac{1}{Z_1}\exp\{Val(i)\}$ for all the candidates $i$, where $Z_1$ is a constant that normalizes the probabilities. Second, we perform $J_0$ runs of this algorithm and average the resulting sparse representations $\{x^{(l)}\}_{l=1}^{J_0}$ that are computed using (\ref{eq:MAP for x}) to obtain the final estimate for $x$. A pseudo-code for the proposed randomized greedy algorithm is given in Algorithm \ref{alg:random greedy MMSE}.

\begin{algorithm}
\caption{Randomized version of Algorithm \ref{alg:greedy MAP} for approximating the MMSE estimator of (\ref{eq:MMSE for x})}
\label{alg:random greedy MMSE}
\begin{algorithmic}
\REQUIRE{Noisy observations $y\in\mathbb{R}^n$, model parameters $W,b,\left\{\sigma_{x,i}\right\}_{i=1}^m,A,\sigma_e$ and number of runs $J_0$.}
\ENSURE{A recovery $\hat{x}_{\textnormal{MMSE}}$ for the representation vector.}
\FOR {$l = 1$ to $J_0$}
\STATE $x^{(l)}=\textbf{0}^{m\times1}$
\STATE $s_*^0=\emptyset,~S_*^0=-\textbf{1}^{m\times1}$
\STATE $k=1$
\REPEAT
\FOR {$i\notin s_{*}^{k-1}$}
\STATE $s^k=s_*^{k-1}\cup i$
\STATE $S^k[j]=\begin{cases} S_*^{k-1}[j] &,\quad j\neq i\\
1 &,\quad j=i\end{cases}$
\STATE Evaluate $Val(i)$ using $(\ref{eq:Val})$.
\ENDFOR
\STATE Choose $i_*$ at random with probabilities $\frac{1}{Z_1}\exp\{Val(i)\}$.
\STATE $s_*^k=s_*^{k-1}\cup i_*,~S_*^k[j]=\begin{cases} S_*^{k-1}[j] &,\quad j\neq i_*\\
1 &,\quad j=i_*\end{cases}$
\STATE $k=k+1$
\UNTIL{$\Pr\left(s_*^k|y\right)<\Pr\left(s_*^{k-1}|y\right)$}
\STATE $\hat{s}=s_{*}^{k-1}$
\STATE Compute $x_{\hat{s}}^{(l)}$ using (\ref{eq:MAP for x}).
\STATE $x_{\hat{s}^C}^{(l)}=0$.
\ENDFOR
\STATE \textbf{Return:} $\hat{x}_{\textnormal{MMSE}}=\frac{1}{J_0}\sum\limits_{l=1}^{J_0} x^{(l)}$
\end{algorithmic}
\end{algorithm}
\vspace{-20pt}

\subsection{Related Pursuit Methods}

\label{sec:related pursuit}

To conclude this section, we mention some related works. First, note that for $W=0$ and equal biases $b_i$ for all $i$, which correspond to an i.i.d. prior, the proposed algorithms resemble the fast Bayesian matching pursuit suggested in \cite{Schniter08}. Second, the recent work of \cite{Cevher08} used a BM-based Bayesian modeling for the sparse representation to improve the CoSaMP algorithm. The inherent differences between our greedy approach and the one suggested in \cite{Cevher08} are explained in Section \ref{sec:past works}.

\section{Exact MAP estimation}

\label{sec:exact MAP}

\subsection{Model Assumptions}

\label{subsec:model assumption for exact MAP}

In this section we consider a simplified setup where exact MAP estimation is feasible. A recent work \cite{Turek11} treated the special case of a unitary dictionary for independent-based priors, and developed closed-form expressions for the MAP and MMSE estimators. We follow a similar route here and assume that the dictionary is unitary. \footnote{In this context we would like to mention that assuming a unitary dictionary $A$ is equivalent to the case $z=x+w$, where there is no dictionary, namely $A$ is the identity matrix, and we have noisy observations of a signal with a BM prior. To see that, multiply each of the sides in the signal equation $y=Ax+e$ by $A^T$. In the resulting equation $A^Ty=x+A^Te$, the noise $w=A^Te$ has the same distribution as the original noise $e$ and $A^Ty$ is the transformed signal. We would like to thank Prof. Phil Schniter for this constructive observation.} In this case we can make a very useful observation which is stated in Theorem \ref{thm:conjugate prior BM}. A proof of this theorem is provided in Appendix \ref{app:A}.

\begin{thm}
Let $A$ be a unitary dictionary. Then the BM distribution is a conjugate prior for the MAP estimation problem of (\ref{eq:MAP for s}), namely the \emph{a posteriori} distribution $\Pr(S|y)$ is a BM with the same interaction matrix $W$ and a modified bias vector $q$ with entries:
\begin{equation}
q_i=b_i+\frac{1}{4}\left\{\frac{\sigma_{x,i}^2}{\sigma_e^2(\sigma_e^2+\sigma_{x,i}^2)}\left(y^Ta_i\right)^2-
\ln\left[1+\frac{\sigma_{x,i}^2}{\sigma_e^2}\right]\right\}
\label{eq:q}
\end{equation}
\noindent for all $i$, where $a_i$ is the $i$th column of $A$.
\label{thm:conjugate prior BM}
\end{thm}

Notice in (\ref{eq:q}) that $q_i$ is linearly dependent on the original bias $b_i$ and quadratically dependent on the inner product between the signal $y$ and the atom $a_i$. This aligns with the simple intuition that an atom is more likely to be used for representing a signal if it has an \textit{a priori} tendency to be turned "on" and if it bears high similarity to the signal (this is expressed by a large inner product). From Theorem \ref{thm:conjugate prior BM} the MAP estimation problem of (\ref{eq:MAP for s}) takes on the form of integer programming. More specifically, this is a Boolean quadratic program (QP):
\begin{equation}
\underset{S}{\textnormal{maximize}}\left(q^TS+\frac{1}{2}S^TWS\right)\textnormal{s.t.}~S_i^2=1,~1\le i\le m.
\label{eq:Boolean QP}
\end{equation}
\noindent This is a well-known combinatorial optimization problem \cite{Du98} that is closely related to multiuser detection in communication systems, a long-studied topic \cite{Verdu98}. The Boolean QP remains computationally intensive if we do not use any approximations or make any additional assumptions regarding the interaction matrix $W$. The vast range of approximation methods used for multiuser detection, like SDP relaxation, can be adapted to our setup. Another approximation approach, which is commonly used for energy minimization in the BM, is based on a Gibbs sampler and simulated annealing techniques \cite{Neal93}. Our interest here is in cases for which simple exact solutions exist. We therefore relax the dependency model, namely make additional modeling assumptions on $W$.

We first consider the simple case of $W=0$, which corresponds to an independency assumption on the entries of $S$. Using Theorem \ref{thm:conjugate prior BM}, we can follow the same analysis as in Section \ref{subsec:Boltzmann} for $W=0$ by replacing the bias vector $b$ by $q$. Consequently, in this case we have:
\begin{equation}
\Pr(S|y)=\prod_{i=1}^m~\Pr(S_i|y),
\label{posterior independence}
\end{equation}
\noindent where $\Pr(S_i=1|y)=\nicefrac{1}{(1+\exp(-2q_i))}$ for all $i$. Notice that $\Pr(S_i=1|y)>\Pr(S_i=-1|y)$ if $q_i>0$. This means that the $i$th entry of $\hat{S}_{_{MAP}}$ equals $1$, namely $i$ is in the support, if $q_i>0$. Using (\ref{eq:q}) we obtain the following MAP estimator for $S$:
\begin{equation}
\hat{S}_{i,MAP}=\left\{\begin{array}{c} 1,\\
-1,\end{array}\right.
\begin{array}{c}
\left|y^Ta_i\right|>\frac{\sqrt{2}\sigma_e}{c_i}\sqrt{\ln\left[\frac{1-p_i}{\sqrt{1-c_i^2}p_i}\right]}\\
otherwise\end{array}
\label{eq:MAP for zero W}
\end{equation}
\noindent where $p_i$ is defined in (\ref{eq:marginals independence}) and $c_i=\sqrt{\nicefrac{\sigma_{x,i}^2}{\left(\sigma_{x,i}^2+\sigma_e^2\right)}}$. These results correspond to those of \cite{Turek11} for the MAP estimator under a unitary dictionary.

To add dependencies into our model, we may consider two approaches, each relying on a different assumption on $W$. First, we can assume that all entries in $W$ are non-negative. If this assumption holds, then the energy function defined by the Boltzmann parameters $W,~q$ is regarded "sub-modular" and it can be minimized via graph cuts \cite{Kolmogorov04}. The basic technique is to construct a specialized graph for the energy function to be minimized such that the minimum cut on the graph also minimizes the energy. The minimum cut, in turn, can be computed by max flow algorithms with complexity which is polynomial in $m$. The recent work \cite{Cevher08} is based on this approach and we will relate to it in more detail in Section \ref{sec:past works}.

Here we take a different approach, which seems to be more appropriate for our setup. This method makes an assumption on the structural component of the MRF - we assume that the BM is \textbf{decomposable with a small tree-width}. This type of MRF was explored in detail in Section \ref{subsec:decomposable models}. The above assumption implies that the matrix $W$ has a special sparse structure - it corresponds to a chordal graph where the size of the largest maximal clique is small. As we have seen in Section \ref{subsec:decomposable models}, decomposable models can serve as a very useful relaxation for general dependency models. Another motivation for this assumption arises from the results that were shown in Section \ref{sec:motivation} for the special case of image patches and a DCT dictionary. It was shown there that independency can be considered a reasonable assumption for many pairs of DCT atoms. This observation has the interpretation of a sparse structure for the interaction matrix $W$. Consequently, it seems plausible that a matrix $W$ with a sparse structure can capture most of the significant interactions in this case.

From Theorem \ref{thm:conjugate prior BM} it follows that if the above assumption on the structure of $W$ holds for the BM prior on $S$ it also holds for the BM posterior (since both distributions correspond to the same interaction matrix). We can therefore use belief propagation techniques to find the MAP solution. We next present in detail a concrete message passing algorithm for obtaining an exact solution to (\ref{eq:Boolean QP}) under a banded $W$ matrix.

To conclude this subsection note that the use of belief propagation techniques \cite{Jordan04} has recently become very popular in the sparse recovery field \cite{Baron10,Donoho10,Schniter10}. However, these works provide a very limited treatment to the structure of the sparsity pattern. We will relate in more detail to these recent works and emphasize the contribution of our work with respect to them in Section \ref{sec:past works}.

\subsection{The Message Passing Algorithm}

\label{subsec:message passing for exact MAP}

Before we go into the details of the proposed message passing algorithm, we make a simple observation that will simplify the formulation of this algorithm. As we have seen in Section \ref{subsec:Boltzmann}, a posterior BM distribution with parameters $W,q$ can be written (up to a normalization factor which has no significance in the MAP estimation problem) as a product of potential functions defined on the maximal cliques in the corresponding graph:
\begin{equation}
\exp\left(q^TS+\frac{1}{2}S^TWS\right)=\prod_{i=1}^{P}\Psi_{C_i}\left(S_{C_i}\right)
\label{eq:product of potentials}
\end{equation}
\noindent where $P$ is the number of maximal cliques. By replacing the potentials $\left\{\Psi_{C_i}\left(S_{C_i}\right)\right\}$ with their logarithms, which are denoted by $\left\{\widetilde{\Psi}_{C_i}\left(S_{C_i}\right)\right\}$, we remain with quadratic functions of the variables of $\left\{S_{i}\right\}_{i=1}^{m}$:
\begin{equation}
S^TWS+q^TS =\sum_{i=1}^{P}\widetilde{\Psi}_{C_i}\left(S_{C_i}\right).
\label{eq:sum of log potentials}
\end{equation}
\noindent This can be very useful from a computational point of view as there is no need to compute exponents, which can lead to large values. Each product that appears in a standard message passing algorithm is replaced by summation.

For concreteness we will focus on the special case of an $L$th order banded interaction matrix $W$ of size $m$-by-$m$, as described in Section \ref{subsec:decomposable models}. In this case the maximal cliques are $C_i=\left\{S_i,\ldots,S_{i+L}\right\},~i=1,\ldots,m-L$, so that all cliques are of size $L+1$ and the tree-width is $L$. The clique tree takes the form of a simple chain of length $m-L$. We denote the "innermost" clique in this chain by $C_k$, where $k=\left\lceil\nicefrac{(m-L-1)}{2}\right\rceil$. We choose an order for the cliques where the cliques at both edges of the chain appear first and the "innermost"  clique appears last and set the clique potentials according to the rule of thumb mentioned in Section \ref{subsec:Boltzmann}. Consequently, the logarithms of the potentials are given by:
\small
\begin{equation}
\widetilde{\Psi}_{C_i}=\begin{cases}
q_iS_i+\sum\limits_{l=i+1}^{i+L}W_{il}S_iS_l & ,~1\leq i\leq k-1\\
\sum\limits_{j=k}^{k+L}q_jS_j+\sum\limits_{j=k}^{k+L-1}\sum\limits_{l=j+1}^{k+L}W_{jl}S_jS_l & ,~i=k\\
q_{i+L}S_{i+L}+\sum\limits_{l=i}^{i+L-1}W_{l,i+L}S_lS_{i+L} & ,~k+1\leq i\leq m-L\end{cases}
\label{eq:clique potentials banded}
\end{equation}
\normalsize
\noindent $\widetilde{\Psi}_{C_i}$ is a function of $S_i,\ldots,S_{i+L}$. We pass messages "inwards" starting from $C_{1}$ and $C_{m-L}$ until the clique $C_{k}$ receives messages from both sides:
\begin{align}
&m_{i,i+1}=\begin{cases}
\underset{S_i}{\max}~\widetilde{\Psi}_{C_{i}} & ,~i=1\\
\underset{S_i}{\max}~\widetilde{\Psi}_{C_{i}}+m_{i-1,i} & ,~2\leq i\leq k-1\end{cases}
\label{eq:messages banded}\\
&m_{i,i-1}=\begin{cases}
\underset{S_{i+L}}{\max}~\widetilde{\Psi}_{C_i} & ,~i=m-L\\
\underset{S_{i+L}}{\max}~\widetilde{\Psi}_{C_i}+m_{i+1,i} & ,~m-L-1\leq i\leq k+1\end{cases}
\notag
\end{align}
\noindent The arguments that correspond to each of the maximization operators are denoted by  $\Phi_{i,i+1},~i=1,\ldots,k-1$ and $\Phi_{i,i-1},~i=k+1,\ldots,m-L$ (these have the same form as the messages with "max" replaced by "argmax"). Note that $m_{i,i+1},\Phi_{i,i+1}$ depend on $S_{i+1},\ldots,S_{i+L}$ and $m_{i,i-1},\Phi_{i,i-1}$ on $S_i,\ldots,S_{i+L-1}$. The MAP estimates are then computed recursively by:
\begin{align}
&\left(S_k^{*},\ldots,S_{k+L}^{*}\right)=\underset{S_k,\ldots,S_{k+L}}
{\textnormal{argmax}}~\widetilde{\Psi}_{C_k}+m_{k-1,k}+m_{k+1,k}\notag\\
&S_i^{*}=\Phi_{i,i+1}\left(S_{i+1}^*,\ldots,S_{i+L}^{*}\right),~i=k-1,\ldots,1\label{eq:MAP banded}\\
&S_{i+L}^{*}=\Phi_{i,i-1}\left(S_i^*,\ldots,S_{i+L-1}^{*}\right),~i=k+1,\ldots,m-L.\notag
\end{align}
\noindent The message passing algorithm in this case is summarized in Algorithm \ref{alg:exact MAP}.

\begin{algorithm}
\caption{Message passing algorithm for obtaining the exact MAP estimator of (\ref{eq:MAP for s})
in the special case of a unitary dictionary and a banded interaction matrix}
\label{alg:exact MAP}
\begin{algorithmic}
\REQUIRE{Noisy observations $y$ and model parameters
$W,b,\left\{\sigma_{x,i}\right\}_{i=1}^m,A,\sigma_e$. $A$ is unitary and $W$ is
an $L$th order banded matrix.}
\ENSURE{A recovery $\hat{S}_{\textnormal{MAP}}$ for the sparsity pattern of $x$.}
\end{algorithmic}
Step 1: Set the bias vector $q$ for the BM posterior distribution $\Pr(S|y)$
using (\ref{eq:q}).\\
Step 2: Assign a potential function $\widetilde{\Psi}_{C_i}\left(S_{C_i}\right)$ for each clique $C_i=\left\{S_i,\ldots,S_{i+L}\right\},~i=1,\ldots,m-L$ using (\ref{eq:clique potentials banded}).\\
Step 3: Pass messages "inwards" starting from $C_{1}$ and $C_{m-L}$ until the "innermost" clique $C_{k}$
receives messages from both sides using (\ref{eq:messages banded}).\\
Step 4: Obtain the MAP estimate for $S$ using (\ref{eq:MAP banded}).
\end{algorithm}

An important observation is that the complexity of the proposed algorithm is exponential in $L$ and not in $m$. More specifically the complexity is $O(2^L\cdot m)$. As the value of $L$ is part of our modeling, even when $m$ is relatively large (and the exhaustive search which depends on $2^{m}$ is clearly infeasible), the exact MAP computation is still feasible as long as $L$ remains sufficiently small. If we have for example $L=\gamma\log_2(m)$ then the complexity is $O(m^{1+\gamma})$, namely it is polynomial in $m$.

\section{Simulations on Synthetic Signals}

\label{sec:synthetic simulations}

In this section we assume that all the parameters of the BM-based generative model are known and use this model to create random data sets of signals, along with their sparse representations. A standard Gibbs sampler is used for sampling sparsity patterns from the BM. The sampled supports and representation vectors are denoted by $\left\{s^{(l)},x^{(l)}\right\}_{l=1}^N$. Using these synthetic data sets, we test the recovery algorithms that were proposed in the two previous sections (see Algorithms \ref{alg:greedy MAP}-\ref{alg:exact MAP}) and compare their performance to that of two previous sparse recovery methods.

The first method is OMP, a standard pursuit algorithm, which serves as the default option that one would use for sparse approximation when no information is given about the structure. The OMP algorithm is used only for identifying the support. Then the recovered support is used to obtain an estimate for the representation vector using (\ref{eq:MAP for x}), just as the MAP estimators. The second is an approximate MAP estimator that is based on Gibbs sampling and simulated annealing as suggested in \cite{Garrigues07}. Since we do not have access to the code and parameters of the algorithm from \cite{Garrigues07}, our implementation is not exactly the same. Rather, we chose to set the number of rounds for the Gibbs sampler so that its computational complexity is roughly the same as the OMP-like MAP method (see Algorithm \ref{alg:greedy MAP}). This choice was made after exploring the influence of the number of rounds for the Gibbs sampler on its performance. We observed that if we increase the number of rounds by a factor of $10$ with respect to the above suggestion, then performance improves only slightly. This behavior is associated with the slow convergence of the Gibbs sampler. As for annealing, we used a geometric schedule: $T_{k+1}=\beta T_k$, where $T_k$ is the "temperature" used in the $k$th round. The initial "temperature" is set to be high (around $600$) and $\beta$ satisfies a final "temperature" of $1$.

We begin by examining a setup that satisfies the simplifying assumptions of Section \ref{sec:exact MAP}. We assume that $A$ is an $m$-by-$m$ unitary DCT dictionary with $m=64$ and that $W$ is a $9$th order banded interaction matrix. The values of the model parameters are in the following ranges: $\left[-1,1\right]$ for the nonzero entries in $W$, $\left[-3,-2\right]$ for the biases $\left\{b_i\right\} _{i=1}^{m}$ and $\left[15,60\right]$ for the variances $\left\{\sigma_{x,i}\right\} _{i=1}^{m}$. In this case we can apply all of the algorithms that were suggested in this paper. However, for concreteness we chose to apply here only Algorithms \ref{alg:greedy MAP},\ref{alg:random greedy MMSE} and \ref{alg:exact MAP}, leaving Algorithm \ref{alg:THR-like MAP} for the second set of synthetic experiments. In Algorithm \ref{alg:random greedy MMSE} we performed $J_0=10$ runs of the random greedy pursuit.

\begin{figure*}
\centering
\includegraphics[scale=0.425]{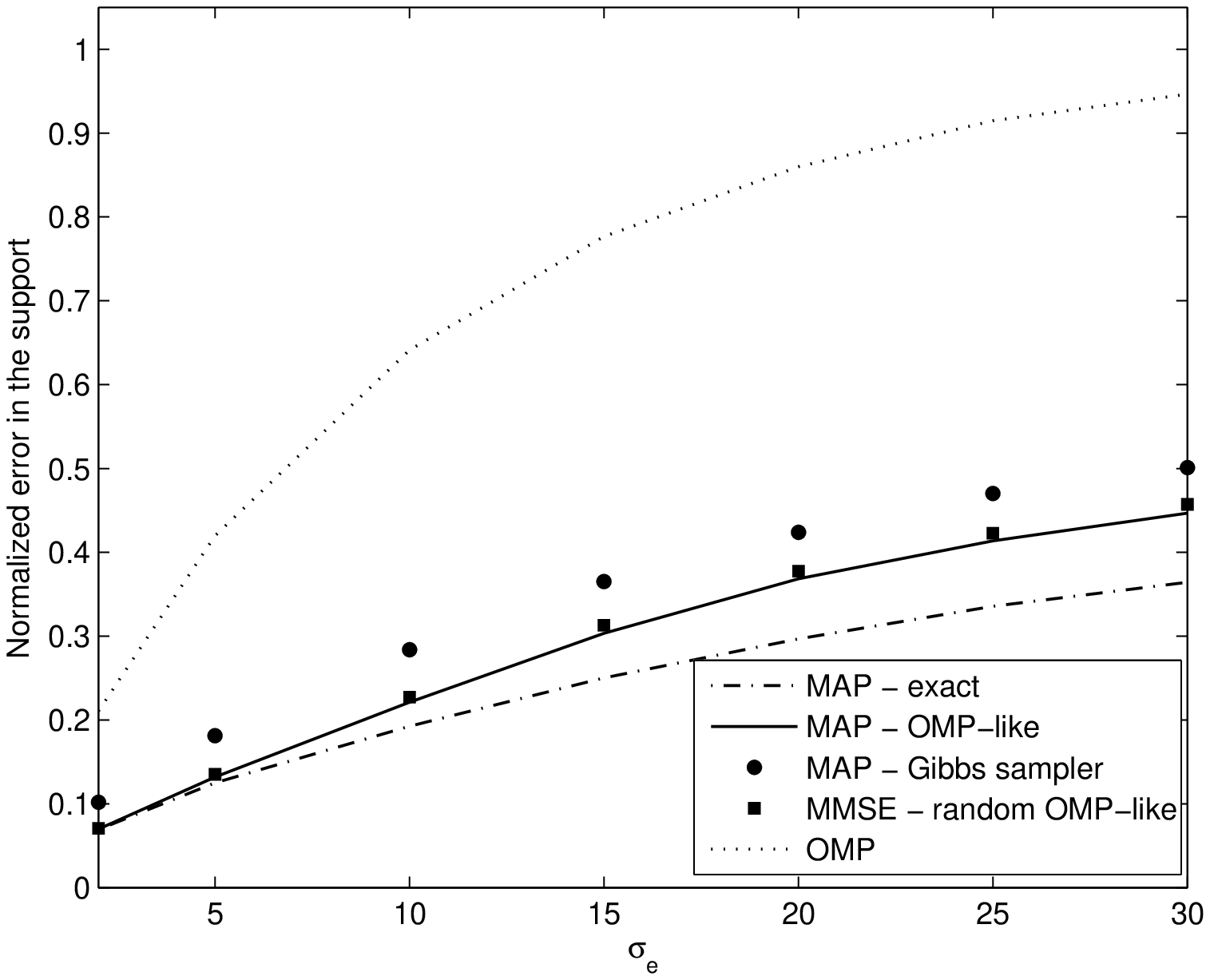}
\includegraphics[scale=0.425]{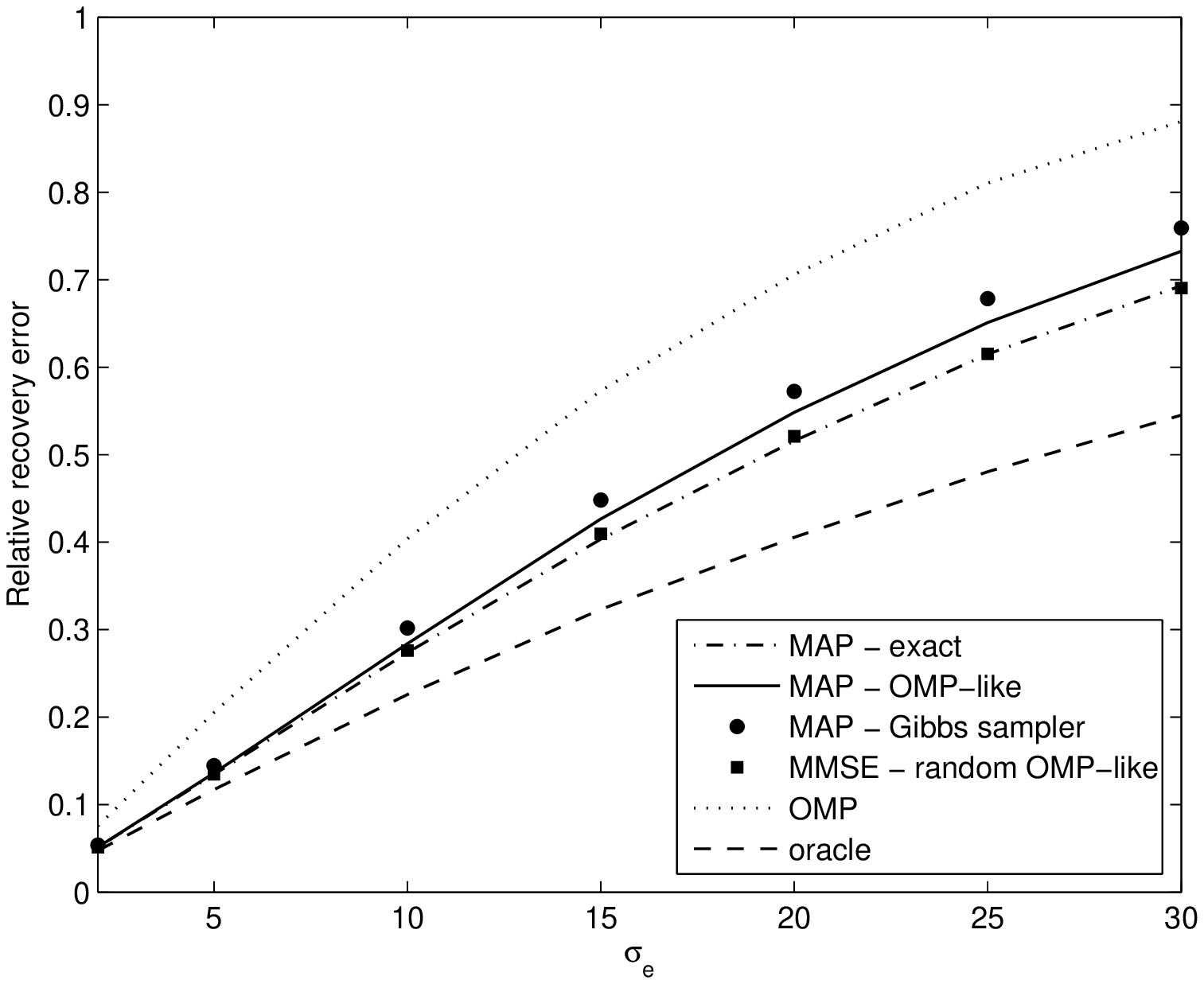}
\caption{Normalized error in identifying the support (\ref{eq:error in support id}) and relative recovery error (\ref{eq:recovery error x}) for the $64$-by-$64$ unitary DCT dictionary and a $9$th order banded interaction matrix. Results are shown for a data set with average cardinality $|s|=9.8$.}
\label{fig:synthetic unitary}
\figminusgap
\end{figure*}

We compare the performance of the five algorithms for different noise levels - $\sigma_{e}$ is in the range $\left[2,30\right]$. For each of the above-mentioned algorithms we evaluate two performance criteria. The first one is the average normalized error in identifying the true support:
\begin{equation}
1-\frac{1}{N}\sum_{l=1}^N\frac{|s^{(l)}\cap\hat{s}^{(l)}|}{\max(|s|,|\hat{s}|)}.
\label{eq:error in support id}
\end{equation}
\noindent Note that for the random greedy algorithm we evaluate the support error using the indices of the $k$ largest coefficients (in absolute value) in the obtained solution $\hat{x}$ as the recovered support $\hat{s}$. The second criterion is the relative recovery error, namely the mean recovery error for the representation coefficients normalized by their energy:
\begin{equation}
\sqrt{\frac{\sum\limits_{l=1}^N\|\hat{x}^{(l)}-x^{(l)}\|_2^2}{\sum\limits_{l=1}^N\|x^{(l)}\|_2^2}}.
\label{eq:recovery error x}
\end{equation}
\noindent The relative error is also evaluated for the Bayesian oracle estimator, namely the oracle which knows the true support. Note that for a unitary dictionary the relative error for the representation coefficients is in fact also the relative error for the noise-free signal, since $\|Au\|_2^2=\|u\|_2^2$ for any vector $u$. The results appear in Fig. \ref{fig:synthetic unitary}.

Several observations can be made from the results in Fig. \ref{fig:synthetic unitary}. First, all BM-based pursuit methods outperform the OMP algorithm. Notice that the message passing algorithm (exact MAP) performs well and the performance of the OMP-like algorithm is not too far off. Second, the OMP-like MAP outperforms Gibbs sampling, for the same computational complexity. Finally, the randomized version of the OMP-like method obtains a recovery error which is roughly the same as exact MAP (recall that the random greedy algorithm approximates the MMSE estimator).

We now provide some additional observations that were drawn from similar sets of experiments which are not shown here. We observed that the performance gaps between the exact MAP and its approximations are associated more with the "strength" of the interactions than with the average cardinality. When we tested higher (less negative) biases and weaker interactions, so that the average cardinality remains roughly the same, the approximations align with the exact MAP (except for the Gibbs sampler which still performs slightly poorer). As for higher noise levels, we noticed that all algorithms exhibit saturation in their performance. In this setup the OMP tends to choose an empty support. The convergence criterion for OMP is $\|y-A_sx_s\|_2<\eta\sqrt{n}\sigma_e$, where $\eta$ is a constant which is close to $1$. This is a standard criterion used for denoising with OMP. When $\sigma_e$ is large, it happens often that the OMP stops before using any atom.

Next, we turn to the case of a redundant dictionary and a general (non-sparse) interaction matrix. We use the $64$-by-$256$ overcomplete DCT dictionary. All the rest of model parameters are the same as before, expect for the interaction matrix which is no longer banded and its values are in the range $[-0.1,0.1]$. For this setup exact MAP estimation is no longer possible and we can use only the greedy approximations for MAP and MMSE (see Algorithms \ref{alg:greedy MAP}-\ref{alg:random greedy MMSE}). We evaluate the average normalized error in the support (\ref{eq:error in support id}) and the relative recovery error with respect to the noise-free signal:
\begin{equation}
\sqrt{\frac{\sum\limits_{l=1}^N\|A\hat{x}^{(l)}-Ax^{(l)}\|_2^2}{\sum\limits_{l=1}^N\|Ax^{(l)}\|_2^2}}.
\label{eq:recovery error y}
\end{equation}

\begin{figure*}
\centering
\includegraphics[scale=0.425]{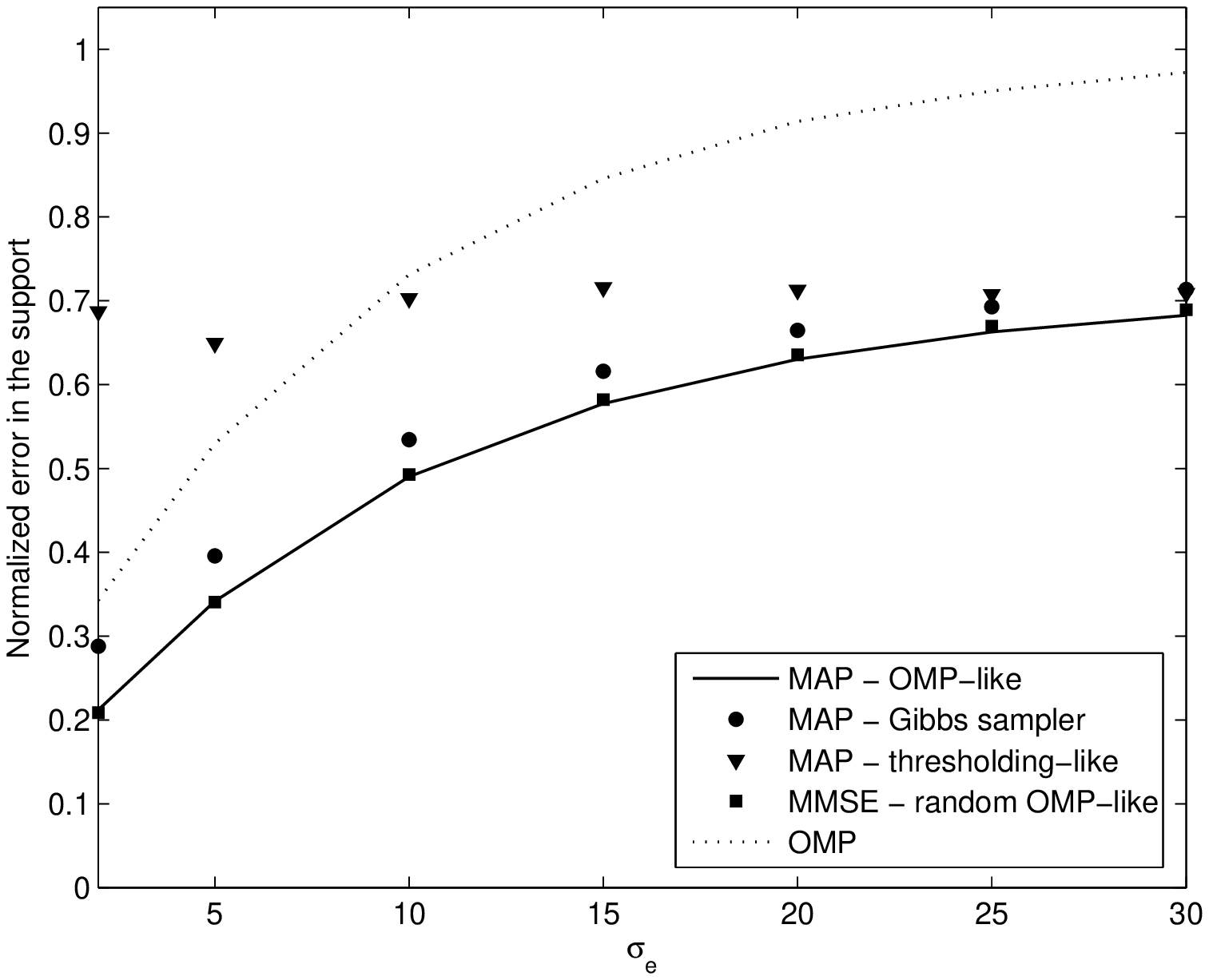}
\includegraphics[scale=0.425]{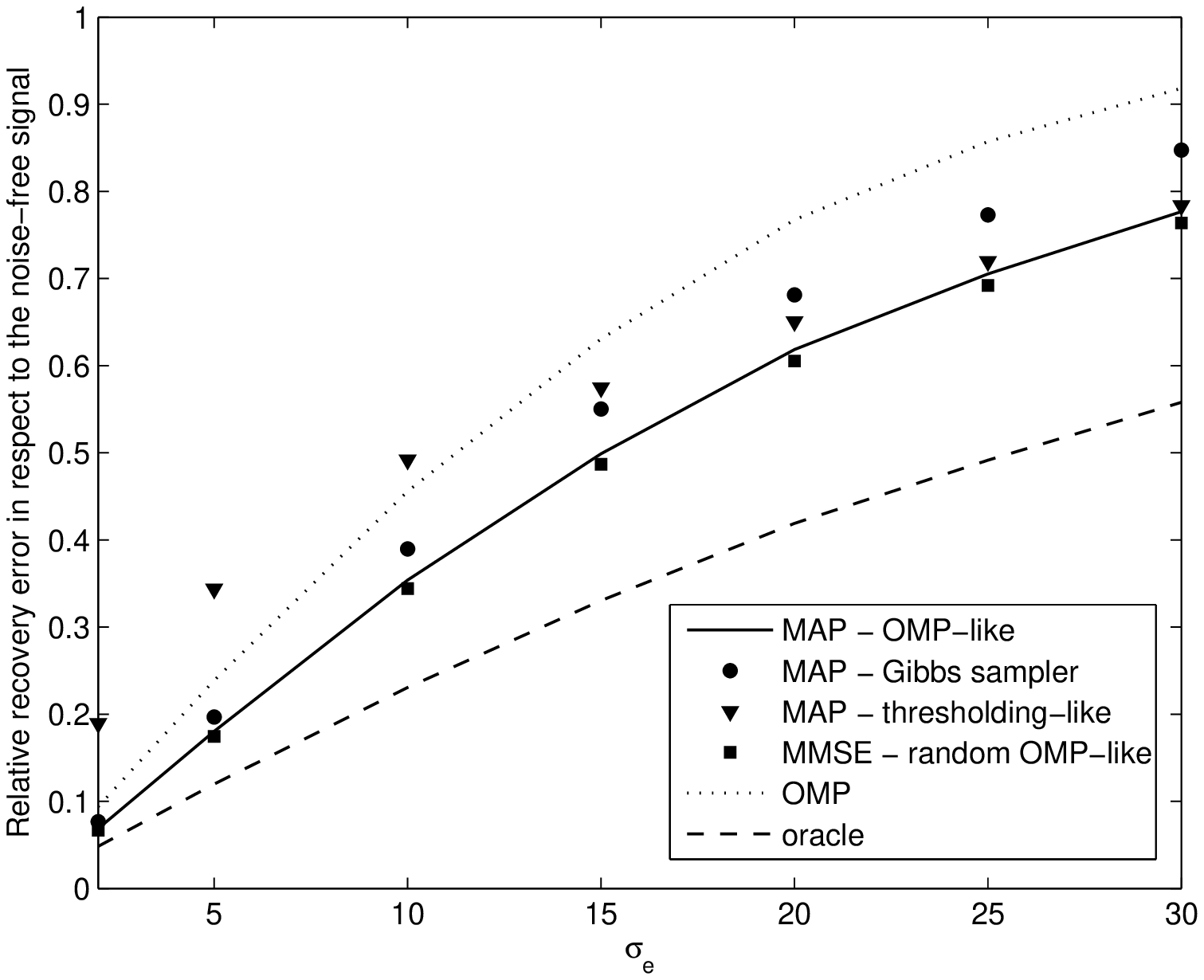}
\caption{Normalized error in identifying the support (\ref{eq:error in support id}) and relative recovery error with respect to the noise-free signal (\ref{eq:recovery error y}) for the $64$-by-$256$ overcomplete DCT dictionary and a general interaction matrix. Results are shown for a data set with average cardinality $|s|=10.3$.}
\label{fig:synthetic overcomplete}
\figminusgap
\end{figure*}

\noindent The results are shown in Fig. \ref{fig:synthetic overcomplete}. We see that both the OMP-like MAP and the Gibbs sampler outperform the OMP algorithm. However, there is a small performance gap in favor of the OMP-like MAP. In terms of the recovery error, we can see that this performance gap increases with the noise level. Notice that the randomized version of the OMP-like method achieves only a slightly better recovery error with respect to the original one. Finally, the thresholding-like method is the worst for noise levels below $\sigma_e=10$ (even OMP performs better). However, as the noise level increases its performance becomes close to that of the OMP-like MAP. Consequently, this method seems adequate for high noise levels.

To conclude this section we discuss the use of the three suggested greedy algorithms, in terms of their computational complexity and recovery quality. The thresholding-like method requires the least computational effort - $O(m)$, but leads to a much inferior recovery, compared to the two other methods. For the OMP-like algorithm the computational complexity is increased by a factor of $k$ - the cardinality of the obtained support - but the recovery improves significantly. The recovery quality can be further improved using the random OMP-like method, however this seems to be a minor improvement given the resulting increase in computations by a factor of $J_0$ ($10$ in our case). In short, the OMP-like method provides the best compromise to the observed tradeoff between computational complexity and recovery quality.

\section{Adaptive Sparse Signal Recovery}

\label{sec:adaptive scheme}

In an actual problem suite we are given a set of signals $\left\{y^{(l)}\right\}_{l=1}^N$ from which we would like to estimate both the sparse representations and the model parameters. We address the joint estimation problem in this section by a block-coordinate relaxation approach. This approach can be applied for both the arbitrary and unitary dictionaries. The only difference is in the pursuit algorithm we use. Note that throughout this section we will assume that the noise variance $\sigma_e^2$ is known. This is a a typical assumption in denoising setups with Gaussian noise. We also assume that the dictionary is fixed and known. Dictionary learning is a common practice in the sparse recovery field (see for example \cite{Aharon06}). However, for concreteness we will not address here how to merge dictionary learning into the adaptive scheme and we leave this for future work.

\subsection{Model Estimation}

\label{subsec:model estimation}

We begin with the model estimation problem. This means that we have a data set of i.i.d. examples $\mathcal{D}=\left\{y^{(l)},x^{(l)},S^{(l)}\right\}_{l=1}^N$, from which we would like to learn the model parameters $\Theta=\left[W,b,\left\{\sigma_{x,i}^2\right\}_{i=1}^m\right]$. To estimate $\Theta$ we suggest a maximum likelihood (ML) approach, and using the BM generative model we can write:
\vspace{-1mm}
\begin{equation}
\hat{\Theta}_{ML}=\underset{\Theta}{\textnormal{argmax}}~\Pr\left(\mathcal{D}|\Theta\right)=
\underset{\Theta}{\textnormal{argmax}}~\sum_{i=1}^m\mathcal{L}(\sigma_{x,i}^2)+\mathcal{L}(W,b),
\label{eq:ML}
\vspace{-2mm}
\end{equation}
\noindent where
\small
\vspace{-1mm}
\begin{equation}
\mathcal{L}(\sigma_{x,i}^2)=\frac{1}{2}\sum_{l=1}^N\left[\frac{1}{\sigma_{x,i}^2}\left(x^{(l)}_i\right)^2+
\ln\left(\sigma_{x,i}^2\right)\right]~\textbf{1}\left[i\in s^{(l)}\right]
\label{eq:log likelihood variances}
\end{equation}
\begin{equation}
\mathcal{L}(W,b)=\frac{1}{2}\sum_{l=1}^{N}~\left[\left(S^{(l)}\right)^TWS^{(l)}+b^TS^{(l)}\right]-N\ln(Z(W,b))
\label{eq:log likelihood Boltzmann}
\end{equation}
\normalsize
\noindent are the log likelihood functions for the model parameters. This decomposition allows separate estimation of the variances $\{\sigma_{x_i}^2\}_{i=1}^m$ and the Boltzmann parameters $W,b$.

Starting with the variances we have the close-form estimator:
\vspace{-1mm}
\begin{equation}
\hat{\sigma}_{x,i}^2=\frac{\sum\limits_{l=1}^{N}\left(x^{(l)}_i\right)^2\textbf{1}\left[i\in s^{(l)}\right]}{\sum\limits_{l=1}^N\textbf{1}\left[i\in s^{(l)}\right]}.
\label{eq:variances estimation}
\end{equation}
\noindent Similar estimators for the variances were also used in \cite{Garrigues07}.

ML estimation of $W,b$ is computationally intensive due to the exponential complexity in $m$ associated with the partition function $Z(W,b)$. Therefore, we turn to approximated ML estimators. A widely used approach is applying Gibbs sampling and mean-field techniques in each iteration of a gradient-based optimization algorithm. These methods were used in \cite{Garrigues07}, which is the only work that considered estimating the BM parameters for sparsity models. However, we suggest using a different approach which seems to be much more efficient - MPL estimation. This approach was presented in \cite{Besag75} and revisited in \cite{Hyvarinen06}, where it was shown that the MPL estimator is consistent. This means that in the limit of infinite sampling ($N\rightarrow\infty$), the PL function is maximized by the true parameter values.

The basic idea in MPL estimation is to replace the BM prior $\Pr(S|W,b)$ by the product of all the conditional distributions of each node $S_i$ given the rest of the nodes $S_{i^{\mathcal{C}}}$: $\prod_{i=1}^{m}\Pr\left(S_i|S_{i^{\mathcal{C}}},W,b\right)$. Each of these conditional distributions takes on the simple form
\begin{equation}
\Pr\left(S_i|S_{i^{\mathcal{C}}},W,b\right)=\tilde{C}\exp\left\{S_i\left(W_i^TS+b_i\right)\right\}
\label{eq:conditional one node BM}
\end{equation}
\noindent where $W_i^T$ is the $i$th row of $W$ and $\tilde{C}$ is a normalization constant. Since this is a probability distribution for a single binary node $S_i$ it follows that $\tilde{C}=\left(2\cosh\left(W_i^TS+b_i\right)\right)^{-1}$. Consequently, we replace $\Pr(S|W,b)$ by
\begin{eqnarray}
&&\prod_{i=1}^{m}\Pr\left(S_i|S_{i^{\mathcal{C}}},W,b\right)=\prod_{i=1}^{m}~
\frac{\exp\left\{S_i\left(W_i^TS+b_i\right)\right\}}{2\cosh\left(W_i^TS+b_i\right)}\notag\\
&&=\frac{\exp\left\{S^T\left(WS+b\right)\right\}}{2^{m}\prod_{i=1}^{m}\cosh\left(W_i^TS+b_i\right)}.
\label{eq:PL Boltzmann}
\eqminusgap
\end{eqnarray}

\noindent We define the log-PL by:
\small
\begin{align}
&\mathcal{L}_p(W,b)=\sum_{l=1}^N\sum_{i=1}^m\ln\left(\Pr\left(S_i^{(l)}|S_{i^{\mathcal{C}}}^{(l)},W,b\right)\right)
\label{eq:log PL Wb}\\
&=\sum_{l=1}^N\left(S^{(l)}\right)^T\left(WS^{(l)}+b\right)-\textbf{1}^T\rho\left(WS^{(l)}+b\right)-mN\ln(2)
\notag
\end{align}
\normalsize
\noindent where $\rho(z)=\ln(\cosh(z))$ and the function $\rho(\cdot)$ operates on a vector entry-wise. To explore the properties of the log-PL function it is useful to place all the Boltzmann parameters - there are $p=\nicefrac{(m^2+m)}{2}$ unknowns ($\nicefrac{(m^2-m)}{2}$ in the upper triangle of $W$ and $m$ in $b$) - in a column vector $u$. For each example $S^{(l)}$ in the data set we can construct matrices $B^{(l)}$, $C^{(l)}$ so that $B^{(l)}u=\left(S^{(l)}\right)^T\left(WS^{(l)}+b\right)$ and $C^{(l)}u=WS^{(l)}+b$.

Using these notations the log-PL function of (\ref{eq:log PL Wb}) can be re-formulated as:
\begin{equation}
\mathcal{L}_p(u)=\sum_{l=1}^N\left[B^{(l)}u-\textbf{1}^T\rho\left(C^{(l)}u\right)\right]-mN\ln(2).
\label{eq:log PL u}
\end{equation}
\noindent The gradient and the hessian of $\mathcal{L}_p(u)$ are given by:
\begin{align}
\nabla\mathcal{L}_p(u)=&\sum_{l=1}^N\left[\left(B^{(l)}\right)^T-\left(C^{(l)}\right)^T
\rho^{\prime}\left(C^{(l)}u\right)\right]\label{eq:grad log PL u}\\
\nabla^2\mathcal{L}_p(u)=&-\sum_{l=1}^N\left[\left(C^{(l)}\right)^T\textnormal{diag}
\left(\rho^{\prime\prime}\left(C^{(l)}u\right)\right)C^{(l)}\right],\label{eq:hessian log PL u}
\end{align}
\noindent where $\rho^{\prime}(z)=\tanh(z)$ and $\rho^{\prime\prime}(z)=1-\tanh^2(z)$. Since $\rho(z)$ is a convex function, it follows that the log-PL function is concave in $u$. Therefore, as an unconstrained convex optimization problem, we have many reliable algorithms that could be of use.

In \cite{Hyvarinen06} MPL estimation is treated by means of gradient ascent (GA) methods. These methods are very simple, but it is well-known that they suffer from a slow convergence rate \cite{Boyd04}. Another optimization algorithm which converges more quickly is Newton \cite{Boyd04}. Note however that the problem dimensions here can be very large. For example, when $m=64$ as in an $8$-by-$8$ image patch and a unitary dictionary, we have $p=2080$ unknown parameters. Since Newton iterations requires inverting the Hessian matrix, it becomes computationally demanding. Instead we would like to use an efficient algorithm that can treat large-scale problems. To this end we suggest the sequential subspace optimization (SESOP) method \cite{Narkiss05}, which is known to lead to a significant speedup with respect to gradient ascent.

\begin{figure*}
\centering
\includegraphics[scale=0.325]{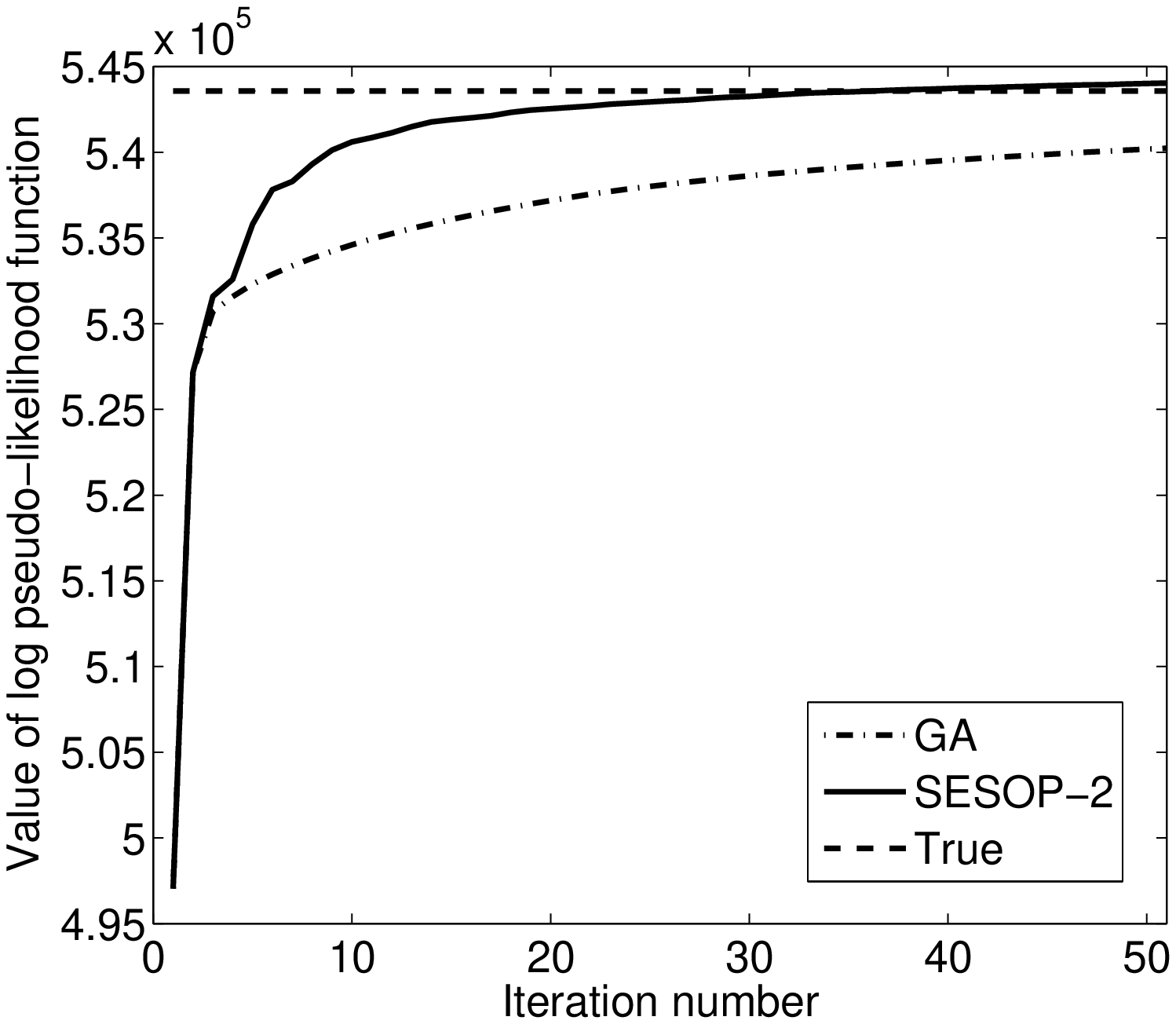}
\includegraphics[scale=0.325]{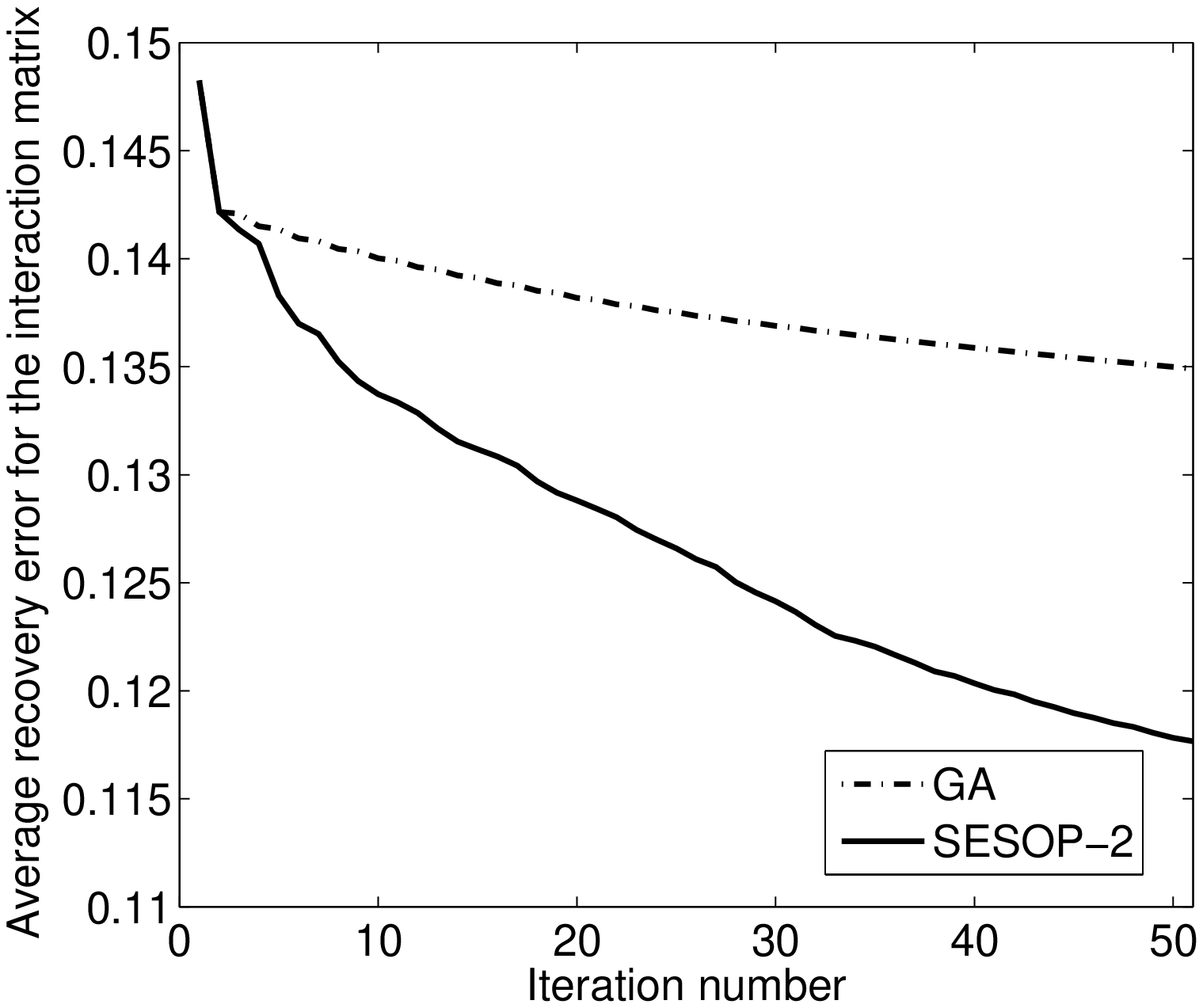}\\
\includegraphics[scale=0.25]{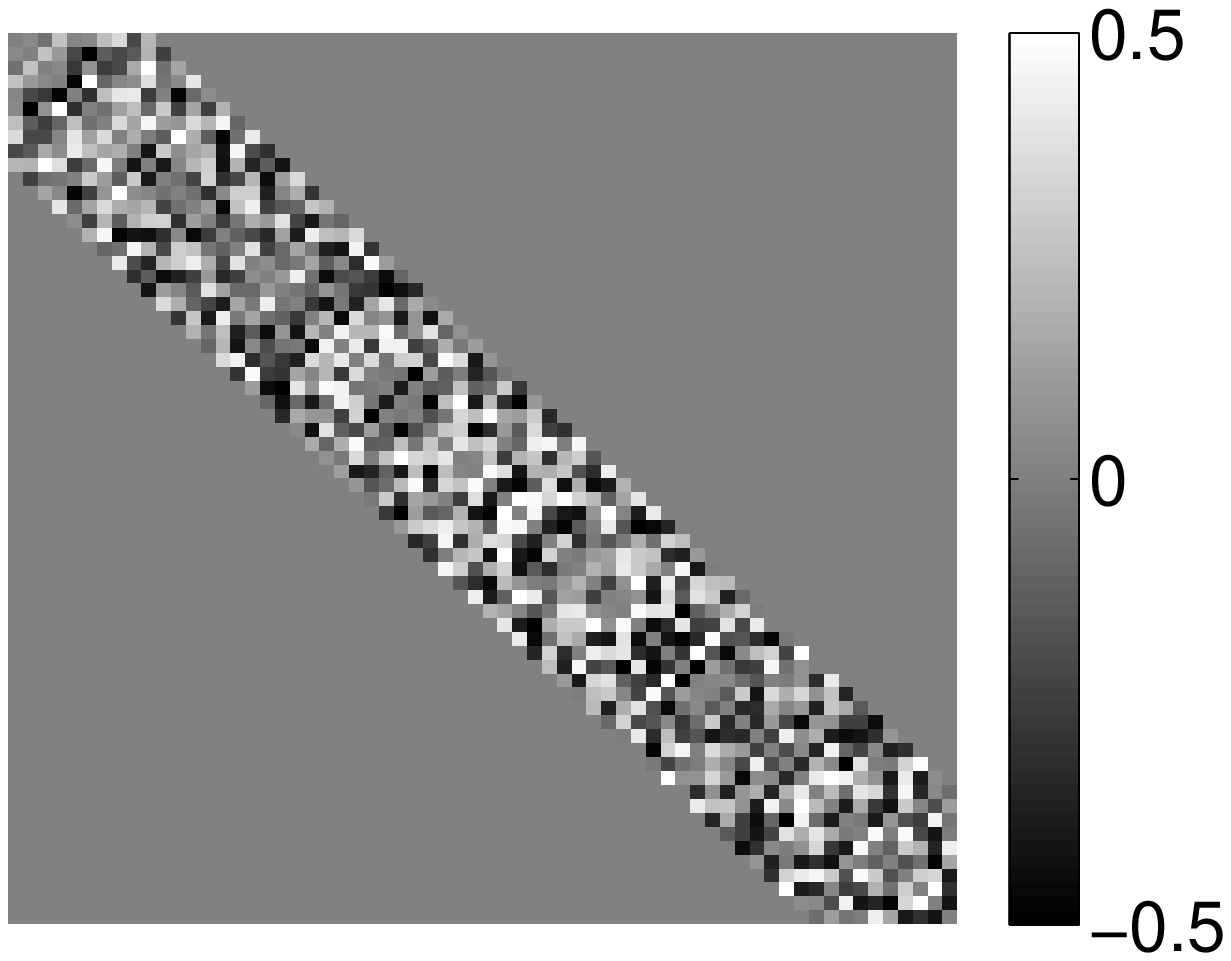}
\includegraphics[scale=0.25]{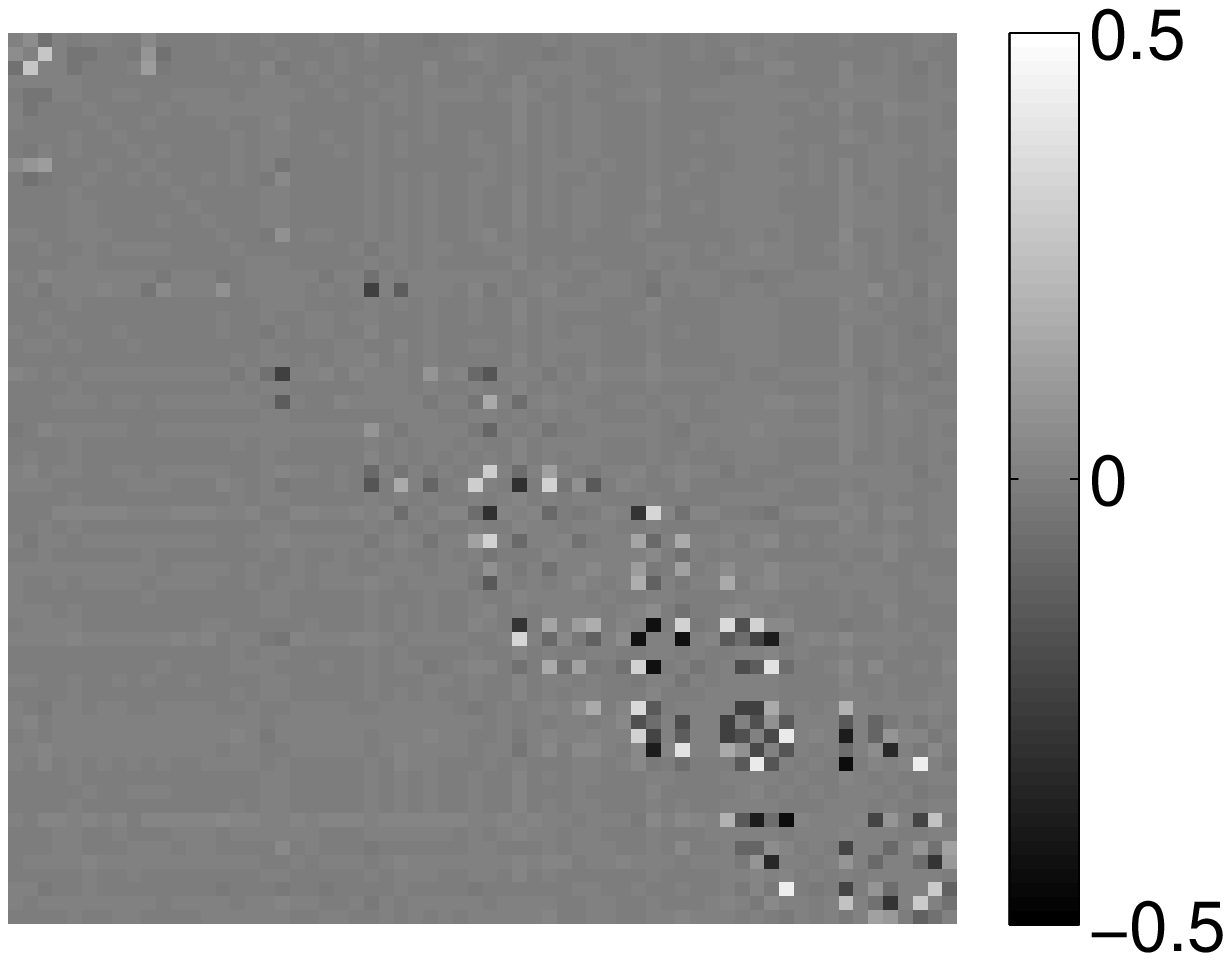}\\
\includegraphics[scale=0.25]{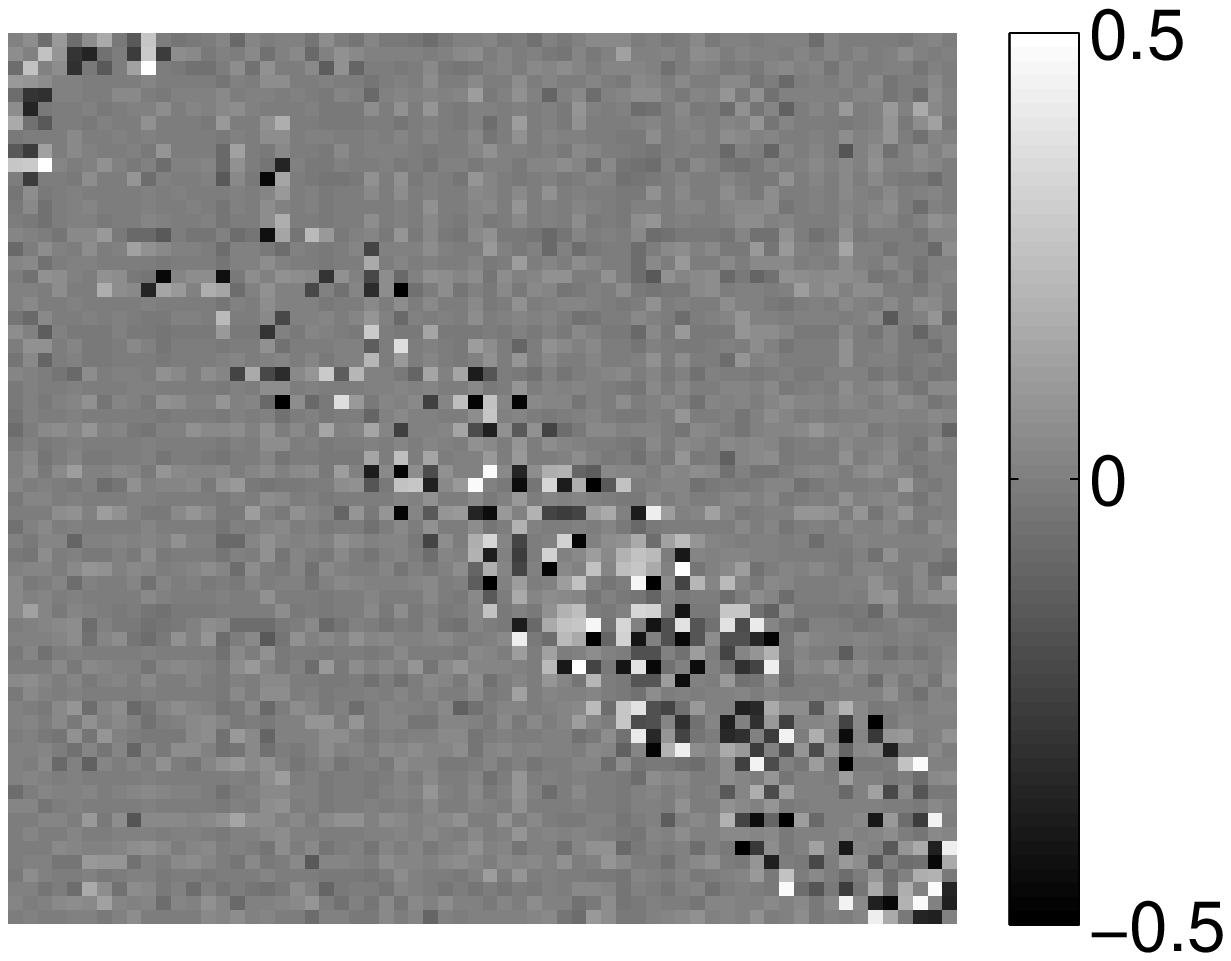}
\includegraphics[scale=0.25]{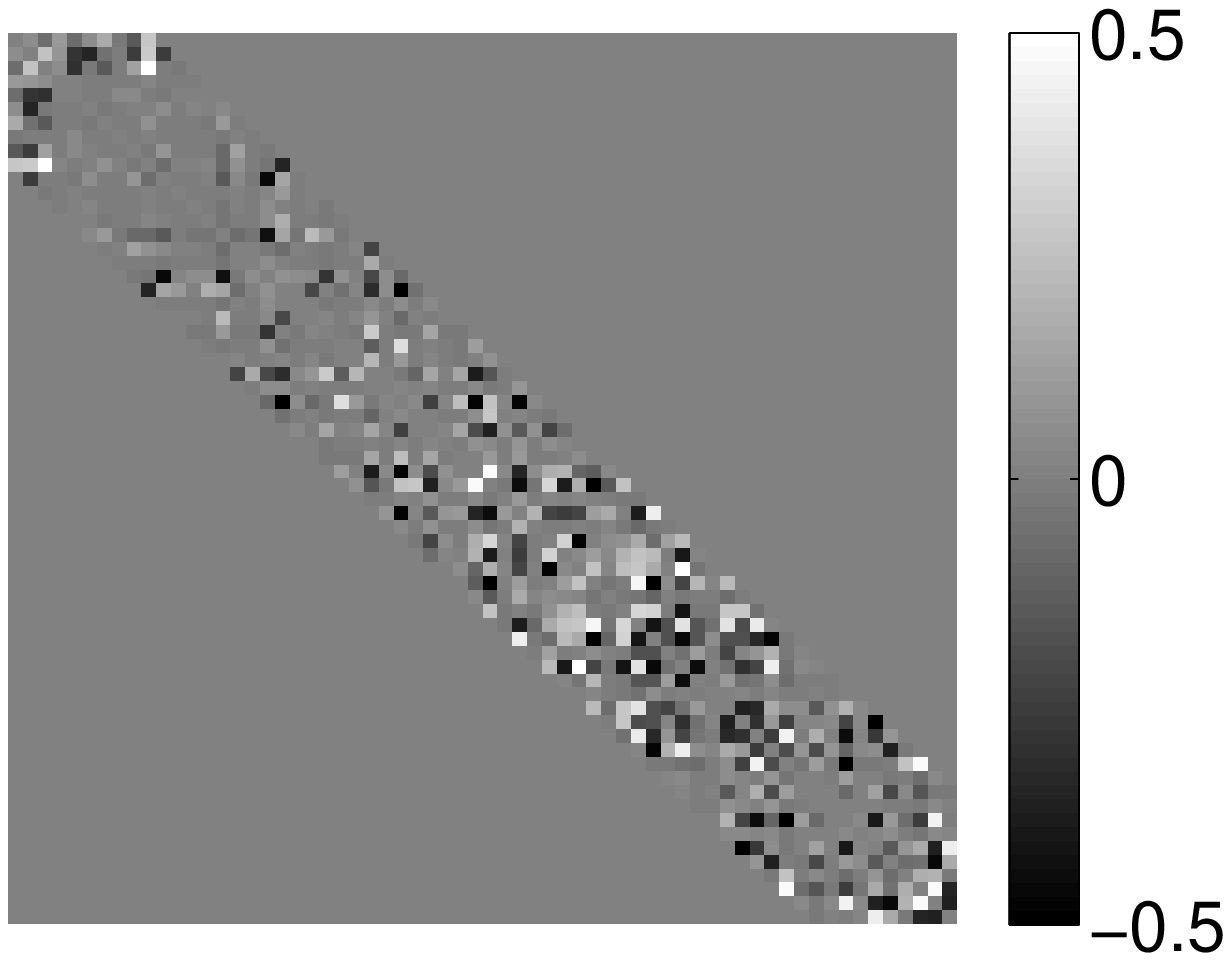}
\includegraphics[scale=0.25]{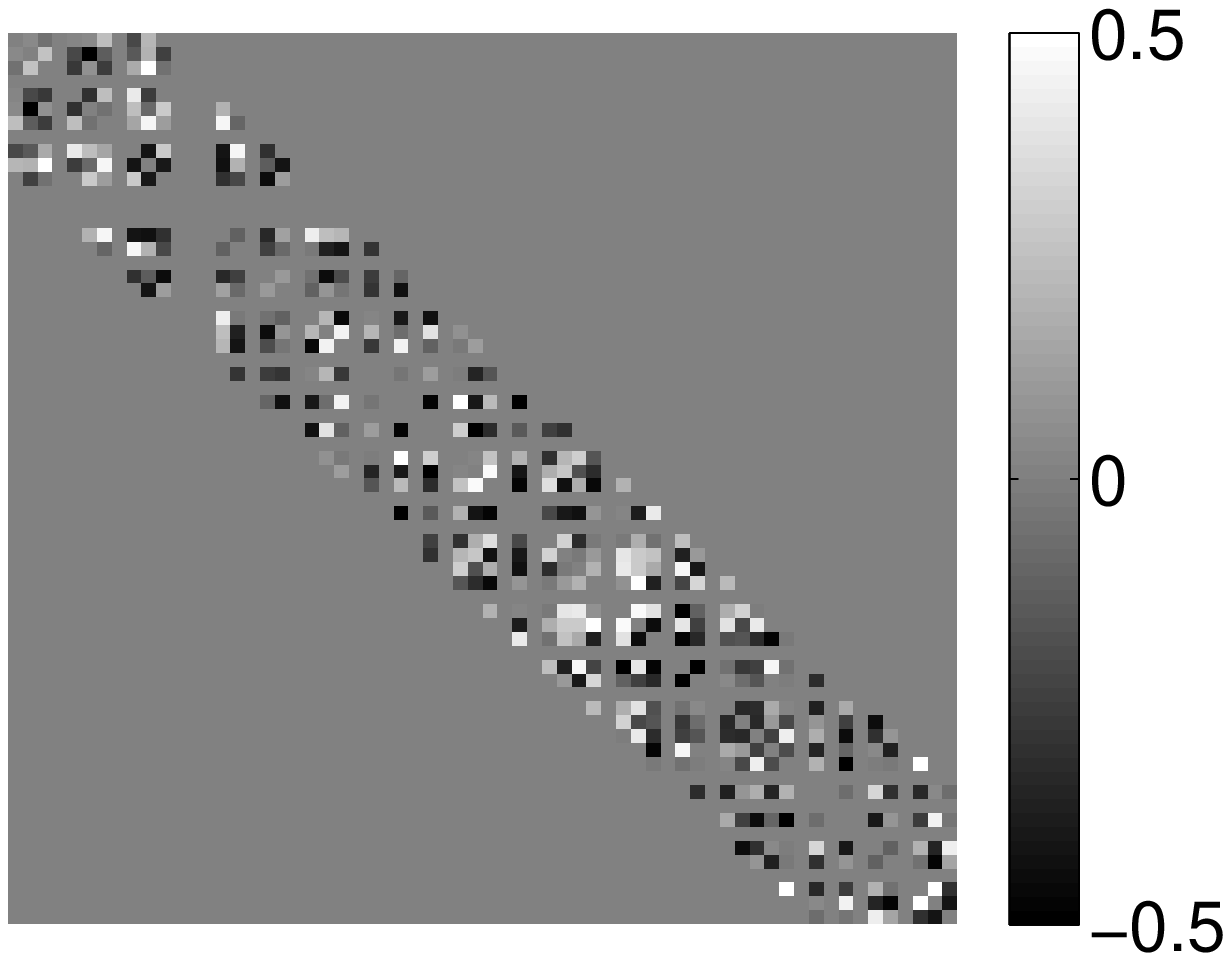}
\caption{Top - results of MPL estimation via GA and SESOP: The value of the log-PL objective and the average recovery error for the interaction matrix per entry as functions of the number of iterations. Middle (from left to right): The true interaction matrix $W$ and MPL estimate via GA $\hat{W}_{GA}$. Bottom (from left to right): MPL estimate via SESOP $\hat{W}_{SESOP}$, a banded version of it and a matrix consisting of the interactions in $W$ which are more likely to be revealed using the given data set. We can see that the latter two are very close.}
\label{fig:MPL results}
\figminusgap
\end{figure*}

The basic idea in SESOP is to use the following update rule for the parameter vector in each iteration:
\begin{equation}
u^{j+1}=u^j+H^j\alpha^j,
\label{eq:update u}
\end{equation}
\noindent where $H^j$ is a matrix consisting of various (normalized) direction vectors in its columns and $\alpha^j$ is a vector containing the step size in each direction. In our setting we use only the current gradient $g^j=\nabla\mathcal{L}_p(u^j)$ and $M$ recent steps $p^i=u^i-u^{i-1},~i=j-M,\ldots,j-1$, so that $H^j$ is a $p$-by-$M+1$ matrix for sufficiently large $j$. We use the abbreviation SESOP-$M$ for this mode of the algorithm. The vector $\alpha^j$ is determined in each iteration by an inner optimization stage. Since we use a small number of directions, maximizing $\mathcal{L}_p(u^{j+1})$ with respect to $\alpha^j$ is a small-scale optimization problem and we can apply Newton iterations to solve it, using $\nabla_{\alpha^j}\mathcal{L}_p(u^{j+1})=\left(H^j\right)^T\nabla\mathcal{L}_p(u^{j+1})$ and $\nabla^2_{\alpha^j}\mathcal{L}_p(u^{j+1})=\left(H^j\right)^T\nabla^2\mathcal{L}_p(u^{j+1})H^j$.

To initialize the algorithm we set the interaction matrix to zero, namely we allow no interactions. We then perform a separate MPL estimation of $b$ where $W$ is fixed to zero, which results in
\begin{equation}
\hat{b}_i^0=\textnormal{atanh}\left[\frac{1}{N}\sum_{l=1}^{N}S_i^{(l)}\right],
\label{eq:initial b estimate}
\end{equation}
\noindent for all $i$. We stop the algorithm either when the norm of the gradient vector $\nabla\mathcal{L}_p(u)$ decreases below a pre-determined threshold $\epsilon$, or after a fixed number of iterations $J_1$. A pseudo-code that summarizes the learning algorithm for the Boltzmann parameters is provided in Algorithm \ref{alg:MPL via SESOP}.

\begin{algorithm}
\caption{A SESOP-$M$ algorithm for obtaining the MPL estimator of the Boltzmann parameters}
\label{alg:MPL via SESOP}
\begin{algorithmic}
\REQUIRE{A data set of supports $\left\{S^{(l)}\right\}_{l=1}^N$.}
\ENSURE{A recovery $\hat{W},\hat{b}$ for the Boltzmann parameters.}
\STATE Initialization: Set $\hat{W}^0$ to zero and $\hat{b}^0$ according to (\ref{eq:initial b estimate}), and construct from them a column vector $\hat{u}^0$.
\STATE $j=0$
\REPEAT
\STATE \textbf{Step 1}: Evaluate $\mathcal{L}_p(\hat{u}^{j})$ and $\nabla\mathcal{L}_p(\hat{u}^{j})$ using (\ref{eq:log PL u})-(\ref{eq:grad log PL u}).
\STATE \textbf{Step 2}: Set the matrix $H^j$ using the current gradient $\nabla\mathcal{L}_p(\hat{u}^{j})$ and $M$ previous steps $\left\{\hat{u}^i-\hat{u}^{i-1}\right\}_{i=j-M}^{j-1}$.
\STATE \textbf{Step 3}: Determine the step sizes $\alpha^j$ by Newton iterations.
\STATE \textbf{Step 4}: $\hat{u}^{j+1}=\hat{u}^j+H^j\alpha^j$.
\STATE $j=j+1$
\UNTIL{$\nabla\mathcal{L}_p(\hat{u}^j)<\epsilon$ or $j\geq J_1$}
\STATE \textbf{Return:} $\hat{W}$, $\hat{b}$ extracted out of $\hat{u}^{j}$.
\end{algorithmic}
\end{algorithm}

To demonstrate the effectiveness of MPL estimation via SESOP, we now show some results of synthetic simulations. We use a Gibbs sampler to generate $N=16,000$ support vectors from a BM prior with the following parameters: $W$ is a $9$th order banded matrix of size $64$-by-$64$ with nonzero entries drawn independently from $\mathcal{U}\left[-0.5,0.5\right]$ and $b$ is a vector of size $64$ with entries drawn independently from $\mathcal{N}\left(-1.5,1\right)$. We then use these supports as an input for the learning algorithm and apply $50$ iterations of both GA and SESOP-$2$ to estimate the Boltzmann parameters. The results are shown in Fig. \ref{fig:MPL results}. We can see on the top that SESOP outperforms GA both in terms of convergence rate of the PL objective and recovery error for the interaction matrix. This is also demonstrated visually on the middle and bottom, where we can see that for the same number of iterations SESOP reveals much more interactions than GA. In fact, if we set to zero the entries in the true $W$ that correspond to rarely used atoms (i.e. if the appearance frequency of atoms $i$ or $j$ is very low then we set $W_{ij}=0$), we can see that SESOP was able to learn most of the significant interactions \footnote{An atom is labeled as "rarely used" if it is active in less than $0.3\%$ of the data samples. This is an arbitrary definition, but it helps in showing that the estimated parameters tend to be close to correct.}.

\subsection{Joint Model Estimation and Pursuit}

\label{subsec:joint estimation}

We now turn to the joint estimation problem, where both the sparse representations and the model parameters are unknown. We suggest using a block-coordinate optimization approach for approximating the solution of the joint estimation problem, which results in an iterative scheme for adaptive sparse signal recovery. Each iteration in this scheme consists of two stages. The first is sparse coding where we apply one of the pursuit algorithms that were proposed throughout this paper to obtain estimates for the sparse representations based on the most recent estimate for the model parameters. If the dictionary is unitary and the interaction matrix is banded we apply the message passing scheme of Algorithm \ref{alg:exact MAP}. Otherwise we use a greedy pursuit (see Algorithms \ref{alg:greedy MAP}-\ref{alg:random greedy MMSE}). This is followed by model update where we re-estimate the model parameters given the current estimate of the sparse representations. We use (\ref{eq:variances estimation}) for the variances and MPL estimation via SESOP (see Algorithm \ref{alg:MPL via SESOP}) for the Boltzmann parameters.

For a setup where the interaction matrix is assumed to be banded, we suggest performing a post-processing of the MPL estimate. More specifically, we define the energy of $W$ as the $\it{l}_1$ norm for the entries in the banding zone. The basic idea is to perform pairwise permutations in $\hat{W}$, namely switch the roles of pairs of atoms, so that the energy will be maximal. A greedy approach can be used, so that in each iteration we replace the roles of one pair of atoms, where this replacement is optimal in the sense of maximizing the energy. The algorithm converges when we cannot increase the energy anymore. At this point we set all entries located outside the banding zone to zero. The suggested post-processing stage serves as a projection onto the banding constraint. Note that the estimated biases and variances should also be modified to account for the changes in the atom roles.

\section{Simulations on Image Patches}

\label{sec:image patches}

\begin{table*}
\begin{center}
\small
\begin{tabular}{||c||c||c||c||c||}
 \hline
 $\sigma_e$ & Unitary OMP & Unitary BM recovery & Overcomplete OMP & Overcomplete BM recovery\\
 \hline
 $2$ & $2.58$ & $\textbf{2.24}$ & $2.52$ & $2.46$\\
 \hline
 $5$ & $4.79$ & $\textbf{4.29}$ & $4.9$ & $4.69$\\
 \hline
 $10$ & $7.55$ & $\textbf{6.85}$ & $7.72$ & $7.19$\\
 \hline
 $15$ & $9.76$ & $\textbf{8.81}$ & $9.94$ & $9.12$\\
 \hline
 $20$ & $11.64$ & $\textbf{10.53}$ & $11.82$ & $10.71$\\
 \hline
 $25$ & $13.33$ & $12.1$ & $13.49$ & $\textbf{12.08}$\\
 \hline
 \hline
\end{tabular}
\end{center}
\caption{Summary of average denoising results (Root-MSE per pixel).}
\vspace{-5mm}
\label{table:denoising results}
\end{table*}

The paper starts with a motivating example on image patches of size $8$-by-$8$ that are extracted out of natural images (see Section \ref{sec:motivation}), showing that there are overlooked dependencies. We now return to this very set of patches and show that the proposed approach does better service to this data. We add white Gaussian noise to these patches and apply the adaptive BM-based sparse recovery scheme that was suggested in the previous section on the noisy patches. We consider two methods that follow this approach. In the first method we fix the dictionary to be the $64$-by-$64$ unitary DCT and assume that the interaction matrix is $9$th order banded. Therefore we use message passing (Algorithm \ref{alg:exact MAP}) for the sparse coding stage and apply post-processing on the learned model parameters to satisfy the banding constraint. The second method uses a fixed overcomplete DCT dictionary of size $64$-by-$256$ and assumes nothing on the interaction matrix. Here we use OMP-like pursuit (Algorithm \ref{alg:greedy MAP}) for sparse coding.

To initialize the parameters of the adaptive BM-based methods, we set all the variances to $50^2$ and use an i.i.d. prior on the support, namely $\Pr(S_i=1)=p$ for all $i$. This prior is obtained by the Boltzmann parameters $\hat{W}=\textbf{0}^{m \times m}$ and $\hat{b}_i=0.5\ln(\nicefrac{p}{(1-p)})$ for all $i$. Note that $p$ has the intuitive meaning of the ratio $\nicefrac{k}{m}$ where $k$ is our prior belief on the mean cardinality of the support. We use a prior belief that the average cardinality for image patches is $k=10$. We then perform two iterations for each of the adaptive schemes.

Note that we are not suggesting here an improved image denoising algorithm, and in contrast to common denoising methods, we do not exploit self-similarities in the image (see for example \cite{Mairal09}). Therefore our comparison is limited to denoising schemes that recover each patch separately. We focus on denoising methods based on the OMP algorithm, since this is the standard pursuit algorithm in patch-based image denoising schemes, see for example \cite{Elad06}. For concreteness we also avoid here comparing our approach with methods that are based on dictionary learning (see for example \cite{Aharon06}). For a comparison with K-SVD denoising \cite{Elad06} which is based on sparse coding via OMP and dictionary learning, see our recent paper \cite{Faktor11}.

We will not show here a comparison to other sturctured sparsity methods, and for a reason. As we mention in Section \ref{sec:intro}, the two most common approaches for structured sparsity are block-sparsity and wavelet trees. These two structures have been successfully incorporated into standard sparse recovery algorithms like CoSaMP, see for example \cite{Baraniuk10}. However, in the image patches experiments of Section \ref{sec:motivation}, the DCT coefficients show neither a tree nor a block-sparsity structure. Therefore in this case it is not even clear how to determine the block structure or tree structure for the pursuit. The alternative is to change the dictionary to a wavelet. Note however that the common practice in wavelet denoising is to apply the wavelet transform on large sub-images, rather than small patches. For this reason we feel that it would not have been fair to compare our results on small patches with the ones obtained by a pursuit method based on wavelet trees.

We compare our approach to two simple denoising schemes which apply the OMP algorithm on the noisy patches using the $64$-by-$64$ unitary DCT and the $64$-by-$256$ overcomplete DCT dictionaries. Throughout this section we use the abbreviations "unitary OMP", "unitary BM recovery", "overcomplete OMP" and "overcomplete BM recovery" to denote the four methods. Average denoising errors per pixel are evaluated for the four methods and for $6$ noise levels: $\sigma_e \in \{2,5,10,15,20,25\}$. A summary of the denoising results is given in Table \ref{table:denoising results}, where the best result for each noise level is highlighted.

These results show that the adaptive BM-based approach suggested throughout this paper obtains better denoising performance on noisy image patches than a standard sparse recovery algorithm such as OMP. For the unitary DCT dictionary, the performance gaps of BM recovery with respect to OMP vary from $0.84[dB]$ to $1.23[dB]$ for the different noise levels. When we turn to the overcomplete DCT dictionary, the performance gaps vary from $0.21[dB]$ to $0.96[dB]$. Note that for both dictionaries OMP obtains a similar performance, with a slight performance gap in favor of the unitary dictionary. As for the BM recovery, the message passing algorithm (used for the unitary case) outperforms the OMP-like algorithm (used for the overcomplete case) for all noise levels, except for $\sigma_e=25$, where the two algorithms exhibit similar performance. This is associated, at least in part, with the accuracy of the pursuit algorithm: exact MAP for the unitary case versus approximate MAP for the overcomplete case. To take full advantage of the redundancy in the dictionary, one should use dictionary learning. We leave this for future work, where we intend to merge dictionary learning into the adaptive scheme, in order to benefit from both the BM generative model and a dictionary which is better fitted to the data.

\section{Relation to past works}

\label{sec:past works}

In this section we briefly review several related works and emphasize the contributions of our paper with respect to them. We begin with recent works \cite{Wolfe04,Garrigues07,Cevher08} that used the BM as a prior on the support of the representation vector. In recent years capturing and exploiting dependencies between dictionary atoms has become a hot topic in the model-based sparse recovery field. In contrast to previous works like \cite{La06,Baraniuk10,Duarte08,He09} which considered dependencies in the form of tree structures, \cite{Wolfe04,Garrigues07,Cevher08} propose a more general model for capturing these dependencies.

The authors of \cite{Wolfe04} use a BM prior on the sparsity pattern of Gabor coefficients to capture persistency in the time-frequency domain. They adopt a non-parametric Bayesian approach and address the estimation problems by MCMC inference methods. In their work the Boltzmann parameters are assumed to be known and fixed. This is contrast to our work where we develop efficient methods for estimating both the sparse representations and the Boltzmann parameters.

The work of \cite{Garrigues07} makes use of a BM prior in the more general context of a sparse coding model, which is represented by a graphical model. They provide a biological motivation for this modeling through the architecture of the visual cortex. We used exactly the same graphical model in our work (see Section \ref{sec:problem formulation}). In \cite{Garrigues07} MAP estimation of the sparse representation is addressed by Gibbs sampling and simulated annealing. These techniques often suffer from a slow convergence rate, so that the algorithm is stopped before the global maximum is reached. In the current work we suggest alternative pursuit methods for MAP estimation. As we have seen in the synthetic simulations of Section \ref{sec:synthetic simulations} our suggested pursuit methods outperform the one suggested in \cite{Garrigues07}.

For learning the Boltzmann parameters the authors of \cite{Garrigues07} suggest Gibbs sampling and mean-field approximations for estimating the gradient of the likelihood function in every iteration of a GA algorithm. In our MPL-based algorithm we suggest a simple update in each iteration, which is based on standard convex optimization methods, instead of the more computationally demanding Gibbs sampling process required in each iteration of the approximate ML algorithm. Our evaluations suggest that there is at least a factor of $10$ in the complexity per iteration, between the Gibbs sampler and a plain GA based on our MPL. Since we have added acceleration (SESOP), the gap between the two methods is in fact even higher, as we will probably need far less iterations.

Next, we turn to \cite{Cevher08}. This work adapts a signal model like the one presented in \cite{Garrigues07}, with several modifications. First, it is assumed that all the weights in the interaction matrix $W$ are nonnegative. Second, the Gaussian distributions for the nonzero representation coefficients are replaced by parametric utility functions. The main contribution of \cite{Cevher08} is using the BM generative model for extending the CoSaMP algorithm, a well known greedy method. The extended algorithm, referred to as lattice matching pursuit (LaMP), differs from CoSaMP in the stage of the support update in each iteration, which becomes more accurate. This stage is now based on graph cuts and this calls for the nonnegativity constraint on the entries of $W$. The rest of the iterative scheme however remains unchanged and is still based on "residuals": in each iteration we compute the residual with respect to the signal and the algorithm stops when the residual error falls below a pre-determined threshold.

There are several fundamental differences between our work and the one reported in \cite{Cevher08}. First, while LaMP exploits the  BM-based generative model only in its support update stage, this model is incorporated into all of the stages of our greedy algorithms, including the stopping rule. Our greedy algorithms work for an arbitrary interaction matrix and in this sense they are more general than LaMP. Furthermore, LaMP requires the desired sparsity level as an input to the algorithm. In contrast, our approach assumes nothing about the cardinality, and instead maximizes the posterior with respect to this unknown. LaMP also makes use of some auxiliary functions that need to be finely tuned in order to obtain good performance. These are hard to obtain for the generative model we are considering. Because of all these reasons, it is hard to suggest a fair experimental comparison between the two works.

We now turn to recent works \cite{Baron10,Donoho10,Schniter10} which considered graphical models and belief propagation for sparse recovery. All of these works represent the sparse recovery setup as a factor graph \cite{Jordan04} and perform sparse decoding via belief propagation. Note however that the first two works use the typical independency assumption on the representation coefficients. More specifically, \cite{Baron10} assumes that the coefficients are i.i.d. with a mixture of Gaussians for their distribution. Hence, the main contribution of these works is exploiting the structure of the observations using graphical models. This is in contrast to our work where we focus on structure in the sparsity pattern in order to exploit dependencies in the representation vector.

The third work \cite{Schniter10} suggests exploiting both the structure of the observations and the structure of the sparsity pattern, using factor graphs and belief propagation techniques. This work is actually more general than ours. However, it leaves the specific problem that we have handled almost untouched. Various structures for the sparsity pattern are mentioned there, including an MRF model. However, the main focus in this paper is how to efficiently combine the observation-structure and pattern-structure. The treatment given for the sparsity-pattern decoding is very limited and empirical results are shown only for a Markov chain structure. This is in contrast to our work where we mainly focus on pattern-structure and address the more general setup of an MRF model.

Finally, our work differs from typical works on model-based sparse recovery, in terms of the signal dimensions. Our work is limited to signals of low-dimensions. This is why we have tested denoising on image patches, where each is of low-dimension. This limitation arises from the fact that, as a general framework, our BM model requires an interaction matrix $W$ which is of size $m$-by$m$. When $m$ is too large (beyond $\sim1000$), estimating $W$ and working with it become a problem. In contrast, tree-based sparse recovery methods, like the ones suggested in \cite{Baraniuk10}, need to work on high-dimensional signals in order to truly benefit from the multi-scale structure of the wavelet coefficients.

\section{Conclusions}

\label{sec:conclusions}

In this work we developed a scheme for adaptive model-based recovery of sparse representations, which takes into account statistical dependencies in the sparsity pattern. To exploit such dependencies we adapted a Bayesian model for signal synthesis, which is based on a Boltzmann machine, and designed specialized optimization methods for the estimation problems that arise from this model. This includes MAP and MMSE estimation of the sparse representation and learning of the model parameters. The main contributions of this work include the development of pursuit algorithms for signal recovery: greedy methods which approximate the MAP and MMSE estimators in the general setup and an efficient message passing algorithm which obtains the exact MAP estimate under additional modeling assumptions. We also addressed learning issues and designed an efficient estimator for the parameters of the graphical model. The algorithmic design is followed by convincing empirical evidence. We provided a comprehensive comparison between the suggested pursuit methods, along with standard sparse recovery algorithms and Gibbs sampling methods. Finally, we demonstrated the effectiveness of our approach through real-life experiments on denoising of image patches. We have released a Matlab toolbox containing all of the suggested BM-based algorithms for both pursuit and model estimation, available at \url{http://www.cs.technion.ac.il/~elad/software/} or \url{http://webee.technion.ac.il/people/YoninaEldar/software.html}.

\begin{appendices}

\section{Proof of Theorem \ref{thm:conjugate prior BM}}

\label{app:A}

We show how the assumption that the dictionary is unitary can be used to simplify the expression for $\Pr(S|y)$.
For a unitary dictionary we have $A_s^TA_s=I$ for any support $s$. Consequently, for a support of cardinality $k$ the matrix $Q_s=A_s^TA_s+\sigma_e^2\Sigma_s^{-1}$ is a diagonal matrix of size $k$-by-$k$ with entries $d_i=1+\nicefrac{\sigma_e^2}{\sigma_{x,i}^2},~i=s_1,\ldots,s_k$ on its main diagonal. Straightforward computations show that the following relations hold:
\begin{align}
y^TA_sQ_s^{-1}A_s^Ty=&\sum_{i\in s}d_i^{-1}(y^Ta_i)^2,\notag\\
\ln\left(\left(\det(Q_s)\right)\right)=&\sum_{i\in s}\ln\left(d_i\right)
\label{eq:unitary relations 1}
\end{align}
\noindent Using the definition of $S$ ($S_i=1$ implies that $i$ is in the support and $S_i=-1$ implies otherwise), we can replace each sum over the entries in the support $\sum_{i\in s}v_i$ by a sum over all possible entries $\sum_{i=1}^{m}\frac{1}{2}\left(S_i+1\right)v_i$. Consequently, the relations in (\ref{eq:unitary relations 1}) can be re-written as:
\begin{align}
y^TA_sQ_s^{-1}A_s^Ty=&\frac{1}{2}\sum_{i=1}^{m}\left(S_i+1\right)d_i^{-1}(y^Ta_i)^2=C_1+\frac{1}{2}f^TS\notag\\
\ln\left(\left(\det(Q_s)\right)\right)=&\frac{1}{2}\sum_{i=1}^{m}\left(S_i+1\right)\ln\left(d_i\right)=
C_2+\frac{1}{2}g^TS
\label{eq:unitary relations 2}
\end{align}
\noindent where $C_1,C_2$ are constants and $f,g$ are vectors with entries $f_i=d_i^{-1}(y^Ta_i)^2$, $g_i=\ln\left(d_i\right)$ for $i=1,\ldots,m$. We now place the relations of (\ref{eq:unitary relations 2}) into the appropriate terms in (\ref{eq:MAP for s}) and get:
\begin{equation}
\ln\left(\Pr(S|y)\right)=C_3+\left(b+\frac{f}{4\sigma_e^2}-\frac{v}{4}-\frac{g}{4}\right)^TS+\frac{1}{2}S^TWS
\label{eq:log posterior}
\end{equation}
\noindent where $C_3$ is a constant. It is now easy to verify that the posterior distribution $\Pr(S|y)$ corresponds to a BM distribution with the same interaction matrix $W$ and a modified bias vector which we denote by $q=b+\frac{f}{4\sigma_e^2}-\frac{v}{4}-\frac{g}{4}$:
\begin{equation}
\Pr(S|y)=\frac{1}{\widetilde{Z}}\exp\left(q^TS+\frac{1}{2}S^TWS\right)
\label{eq:posterior}
\end{equation}
\noindent where $\widetilde{Z}$ is a partition function of the BM parameters $W,q$ which normalizes the distribution. Using the definitions of $f$, $g$ and $v$ we get that (\ref{eq:q}) holds.
\end{appendices}

\bibliography{references}
\bibliographystyle{IEEEbib}

\end{document}